\documentclass{aa}

\usepackage{graphicx}
\usepackage{txfonts}

\usepackage{blindtext}
\usepackage{hyperref}
\usepackage{rotating}
\usepackage{graphicx,capt-of}
\usepackage{longtable}
\usepackage{comment}
\usepackage{afterpage}
\usepackage{siunitx}
\usepackage{float}
\usepackage{hyperref}
\usepackage{multirow}

\usepackage{natbib}

\hypersetup{%
        colorlinks=true, linkcolor=blue, 
        breaklinks,
        citecolor = blue,
        unicode=true,
	      }

\begin{document} 

   \title{Surface brightness--colour relations of Milky Way and Magellanic Clouds classical Cepheids based on Gaia magnitudes}

   \titlerunning{SBCR of MW and MC Cepheids based on Gaia magnitudes}
   \authorrunning{M.C. Bailleul et al.}

   \author{M.C. Bailleul
          \inst{1}
          \and
          N. Nardetto
          \inst{1}
          \and
          V. Hocdé
          \inst{1}
          \and
          P. Kervella
          \inst{3,4}
          \and W. Gieren \inst{5}
          \and J. Storm \inst{6}
          \and G. Pietrzyński \inst{2}
          \and A. Gallenne \inst{7}
          \and \\ 
          D. Graczyk \inst{2}
          \and G. Bras \inst{3}
          \and O. Creevey \inst{1}
          \and A. Recio Blanco \inst{1}
          \and P. de Laverny \inst{1}
          \and P.A. Palicio \inst{1}
          \and W. Kiviaho \inst{3}
          }

   \institute{
             Université Côte d’Azur, Observatoire de la Côte d’Azur, CNRS, Laboratoire Lagrange, Nice, France \\
             \email{manon.bailleul@oca.eu}
         \and
             Nicolaus Copernicus Astronomical Center, Polish Academy of Sciences, ul. Bartycka 18, 00-716 Warszawa, Poland
         \and
            LIRA, Observatoire de Paris, Universit\'e PSL, Sorbonne Universit\'e, Universit\'e Paris Cit\'e, CY Cergy Paris Universit\'e, CNRS, 5 place Jules Janssen, 92195 Meudon, France
         \and
            French-Chilean Laboratory for Astronomy, IRL 3386, CNRS, Casilla 36-D, Santiago, Chile
         \and
         Universidad de Concepción, Departamento Astronomía, Casilla 160-C, Concepción, Chile
        \and
            Leibniz Institute for Astrophysics, An der Sternwarte 16, 14482 Potsdam, Germany
        \and
            Instituto de Alta Investigaci\'on, Universidad de Tarapac\'a, Casilla 7D, Arica, Chile
             }

   \date{}

 
  \abstract
   {Surface brightness–colour relations (SBCRs) are widely used to estimate the angular diameters of stars. In particular, they are employed in the Baade–Wesselink distance determination method, which relies on the comparison between the linear and angular amplitudes of Cepheids. The SBCR can be calibrated by combining different photometric systems. An SBCR was recently calibrated based on Gaia DR3 magnitudes alone for fundamental-mode classical Cepheids with solar metallicity. This relation appears to be strongly affected by metallicity, however.}
   {We derive SBCRs for classical Cepheids in the Milky Way and in the Magellanic Clouds using the photometric data available in the \textit{Gaia} database, and we quantify the metallicity effect.}
   {We first selected the data on the basis of a number of quality criteria and chose the best photometric data and the best parallaxes available in \textit{Gaia} for Milky Way classical Cepheids. Secondly, we compiled an extensive list of period-radius (PR) relations available in the literature, and we also provide a new PR relation based on interferometric data in our previous work. Thirdly, combining the radius of classical Cepheids with distance estimates (based on Gaia parallaxes for the Milky Way and on eclipsing binaries for the Magellanic Clouds), we derived the surface brightness and colour of about 1700 classical Cepheids.}
   {We first derived a new PR relation based on interferometric data and distances from the literature of seven classical Cepheids: $\mathrm{\log(R/R_{\odot}) = 1.133_{\pm 0.019} + 0.688_{\pm 0.016} log(P)}$. The metallicity does  not affect the PR relations. Secondly, we calculated three different SBCRs for the Milky Way and Large and Small Magellanic Cloud classical Cepheids based on this new PR relation that clearly show the dependence of the metallicity on the SBCR based on \textit{Gaia} magnitudes alone. Finally, we derived relations between the slopes, the zero points (ZP), and the metallicity ([Fe/H]) of these three SBCRs: $\mathrm{Slope_{SBCR}=-0.0663_{\pm 0.0121} [Fe/H] - 0.3010_{\pm 0.0030}}$ and $\mathrm{ZP_{SBCR}=-0.1016_{\pm 0.0091} [Fe/H] + 3.9988_{\pm 0.0029}}$.}
   {These new SBCRs, dedicated to classical Cepheids in the Milky Way and Magellanic Clouds, are of particular importance to apply the inverse Baade-Wesselink method to classical Cepheids observed by Gaia in a forthcoming study.}

 \maketitle 

%________________________________________________________________

\section{Introduction}
\label{sec:introduction}
The surface brightness--colour relation (SBCR) is commonly used to determine the angular diameter of stars based on their photometric measurements in two different bands. These relations are used, for example, in studies of exoplanet host stars \citep{gent_2022, di_mauro_2022} or asteroseismic targets \citep{valle_2024, campante_2019}. They were also used to derive the distance of eclipsing binaries in the Large \citep{pietrzynski_2013, pietrzynski_2019} and Small Magellanic Clouds \citep{graczyk_2020} with an unprecedented accuracy of 1 and 2\%, respectively, which is of particular importance for the determination of the Hubble-Lemaitre constant \citep{riess_2022, riess_2024, CosmoVerse_valentino_2025}. The SBCR is also used in the context of the Baade-Wesselink (BW) method of Cepheid distance determinations \citep{wesselink_1969}. The BW method (or parallax of pulsation method) compares the linear and angular dimensions of Cepheids to derive its distance \citep{kervella_2004_I_distances, gieren_2018, merand_2015_spips}. \cite{salsi_2022} has shown theoretically based on atmospheric models that SBCRs depend not only on the temperature, but also on their luminosity class. When the BW method is applied to Cepheids, an SBCR dedicated to Cepheids is therefore required, in particular because the BW method is very sensitive to the choice of the SBCR \citep{nardetto_2023_pfactor}. 
One of the largest current photometric survey of Cepheids is the \textit{Gaia} survey, which has observed thousands of Cepheids in three different passbands \citep{theGaiaMission_2016, chapterVariability_2022gdr3}. Recently, \cite{bailleul_2025} have calibrated a new SBCR based on \textit{Gaia} bands alone that is dedicated to classical Cepheids (hereafter, Cepheids refers to fundamental-mode classical Cepheids), which opens the road to the application of the BW method to thousands of Cepheids in the \textit{Gaia} database. We also found in this study that the SBCR is highly sensitive to metallicity, however, which limits its application.

In this work, we calibrate the SBCR in the Gaia photometric bands $\mathrm{G_{BP}}$ and $\mathrm{G_{RP}}$ for Milky Way (MW) and Magellanic Cloud (MC) Cepheids for the first time using an inverse method in which the radius is estimated from a period-radius (PR) relation, while the distance is based on Gaia parallaxes for the MW Cepheids and eclipsing binaries for the MC Cepheids.
The paper is structured as follows. In Sect.~\ref{sec:data_selection} we present the selection process on the \textit{Gaia} data. In Sect.~\ref{sec:period_radius_relations} we present a compilation of PR relations and a new PR relation derived from interferometric measurements and various distance estimates that is then specifically used to calibrate the SBCR. In Sect.~\ref{sec:distances} we discuss the corrections applied to the \textit{Gaia} parallaxes and the choice of distances to the Magellanic clouds. We then present the method we applied to the photometric data from \textit{Gaia} and the extinction we used to correct for them in the sections \ref{sec:surface_brightness_colour_relation} and \ref{sec:extinction}. We discuss the metallicity of the MW and MC Cepheids in Sect.~\ref{sec:metallicity}. Finally, we present in Sect.~\ref{sec:results} the SBCRs we found for MW and the MC Cepheids and conclude.

\section{Data selection}
\label{sec:data_selection}
\begin{figure}[h!]
    \centering
    \includegraphics[width=0.9\linewidth]{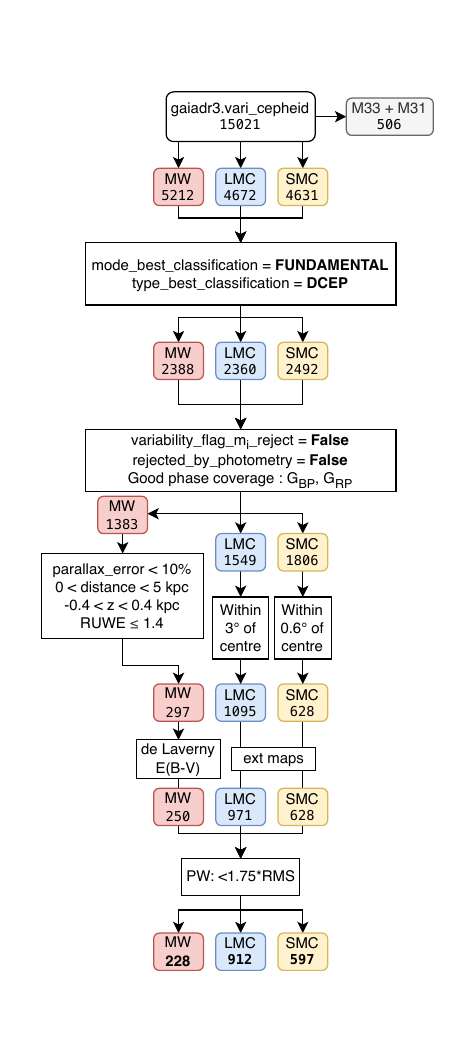}
    \caption{Data selection process.}
    \label{fig:data_selection}
\end{figure}

\begin{figure*}[h!]
    \centering
    \includegraphics[width=1\linewidth]{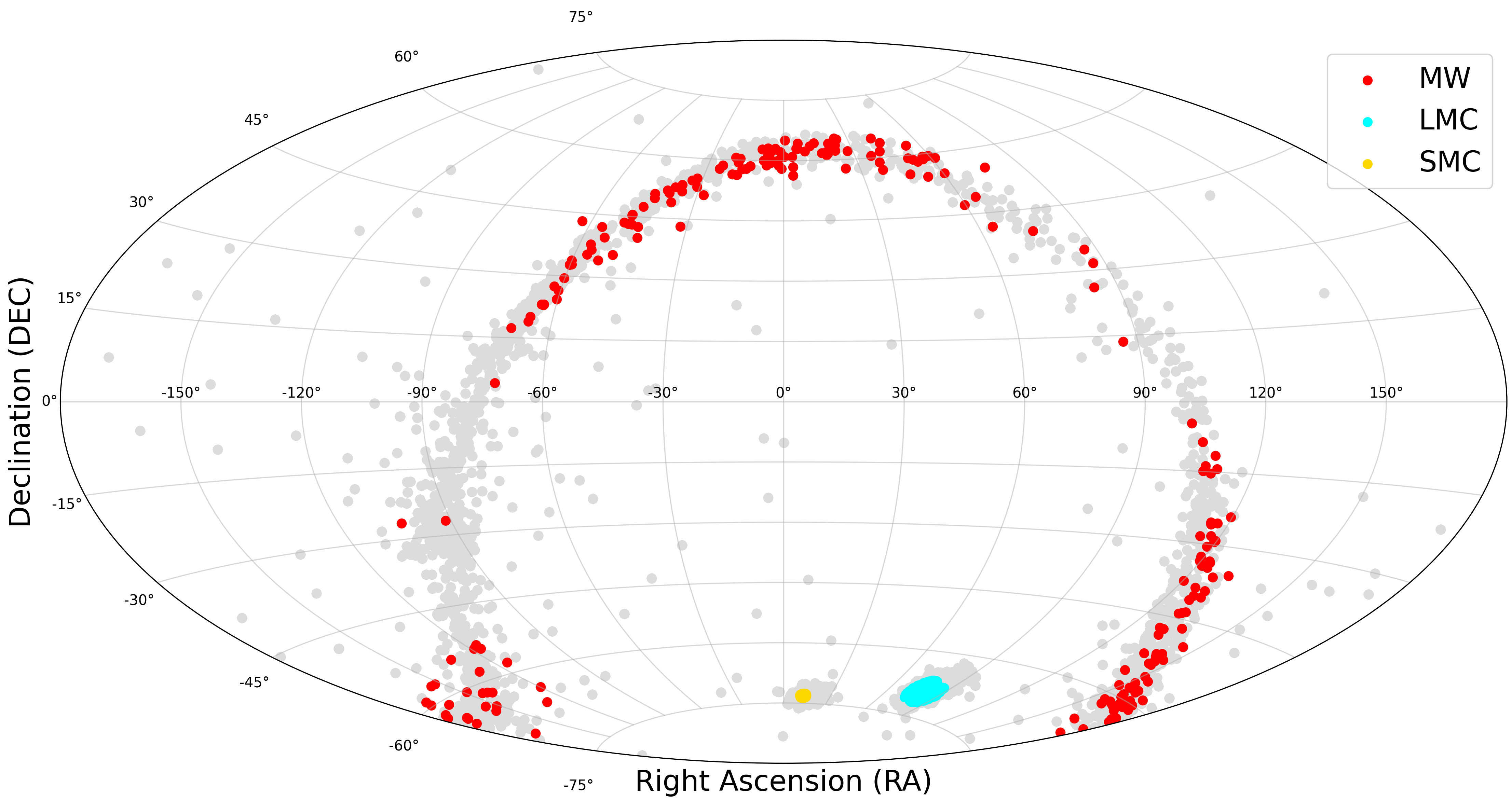}
    \caption{Gaia all-sky view of the Milky Way and the Large and Small Magellanic Clouds. We show the total sample in grey. The final selected sample of Cepheids is shown in colour.}
    \label{fig:gaia_all_sky_view}
\end{figure*}

We retrieved the light curves in the $\mathrm{G_{BP}}$ and $\mathrm{G_{RP}}$ magnitude bandpasses of all Cepheids from the Gaia DR3 database, as well as some specific parameters \citep{gaia_collab_2022_variability, gaia_collaboration_gaia_2023_phot, gaia_collab_vallenari_2023_instru}. We are particularly interested in the following parameters: type, mode, parallax, parallax error, period, and the renormalised unit weight error (RUWE).
The selection process is described in Fig.~\ref{fig:data_selection}.
We first selected the DCEP (corresponds to classical Cepheids in the \textit{Gaia} classification) and FUNDAMENTAL Cepheids. This first study is devoted to classical Cepheids alone because they have been extensively observed by interferometry in the past decades, which is not the case for first-overtone or anomalous Cepheids. This helped us to first apply our method consistently because we used a PR relation based on interferometry (see Sect.~\ref{sec:period_radius_relations}), and it also allowed us to compare our result with a direct recent calibration of the SBCR by interferometry (see Sect.~\ref{sec:results}). Secondly, we rejected poor-quality photometry using three flags: \texttt{variability\_flag\_bp\_reject}, \texttt{variability\_flag\_rp\_reject},~\texttt{rejected\_by\_photometry}. Third, we selected only stars with a good phase coverage, which means at least one measurement per bin in 0.1 in phase. This minimum coverage ensured that we derived a robust mean magnitude (see Sect.~\ref{sec:surface_brightness_colour_relation}). %For the Milky Way's stars, we applied a special selection to the parallax measurement, as it will be useful in deriving the distances of these stars. 
To derive the pulsation phase, we used the following method: We phased the data using the observation time given by the catalogue for each band (i.e. \texttt{bp\_obs\_time} and \texttt{rp\_obs\_time}) and using $\mathrm{\phi_{i} = \frac{JD_{i}-T_{0}}{P_{ref}} \pmod{1}}$, where $\mathrm{T_{0}}$ is the reference epoch, and $\mathrm{P_{ref}}$ is the pulsation period.
Fourth, we selected the stars whose relative parallax error was smaller than 10\%. To be consistent with the grid of the G-Tomo extinction maps discussed in Sect.~\ref{sec:extinction}, we removed Cepheids whose distances were smaller than 5kpc and whose absolute height ($z$) relative to the Galactic plane exceeded 400 pc. This selection focused on Galactic Cepheids located in the plane in the Solar vicinity. We note that this selection biased our Cepheid sample towards a shorter pulsation period since their long-period counterparts are mostly found closer to the Galactic centre beyond 5kpc. For the MW Cepheids, our period ranged between 2 and 45 days, with a median of 5.7 days. In addition, we set an upper limit on the RUWE indicator to 1.4. This parameter gives information about the quality of the fit of the astrometric observations, and a value higher than 1.4 might indicate that the source is not single or that there is a problem for the astrometric solution \citep{lindegren_2021one}. 
Fifth, we selected Large Magellanic Cloud (LMC) and Small Magellanic Cloud (SMC) stars within a radius of 3° and 0.6°, respectively, around the centre of the clouds in order to limit the distance dispersion of these stars (see Sect.~\ref{sec:distances}). 
Sixth, some stars were removed according to the limits of the extinctions maps we used and the availability of extinction values (see Sect.~\ref{sec:extinction}).
Finally, we rejected all the Cepheids that did not follow the period-Wesenheit (PW) relation as described in Sect.~\ref{sec:results}.
At the end, we obtain 9.5\%, 38.6\%, and 24.0\% of the total sample of DCEP and FUNDAMENTAL Cepheids available in the \textit{Gaia} database for the Milky Way, LMC, and SMC, respectively. The whole selected sample is also visible in Fig.~\ref{fig:gaia_all_sky_view} in terms of right ascension and declination.

\section{Period-radius relations}
In this section, we compile the PR relations available in the literature for Cepheids (see Tab.~\ref{tab:PR_relations}) based on the BW and CORS methods and on theory. In \ref{subsec: The BW method+pr} we derive a new relation that is purely observational and is based on the interferometric dataset provided by \cite{bailleul_2025}.
\label{sec:period_radius_relations}

\begin{table*}[ht]
\caption{Period-radius relations from the literature.}
\centering
\begin{tabular}{|l|l|l|l|}
        \hline
        Reference & Method & Expression & Valid For \\
        \hline
        \cite{fernie_survey_1984} & BW, compilation & $\mathrm{1.240_{\pm0.022} + 0.589_{\pm0.022} \log P}$ & MW \\
        \cite{moffett_observational_1987} & BW, $p$ cst, SBCR & $\mathrm{1.110_{\pm0.033} + 0.740_{\pm0.035} \log P}$ & MW \\
        \cite{moffett_observational_1987} & BW, $p$ var, SBCR & $\mathrm{1.131_{\pm0.033} + 0.734_{\pm0.034} \log P}$ & MW \\
        \cite{rojo_arellano_new_1994} & BW, SBCR & $\mathrm{1.125_{\pm0.030} + 0.744_{\pm0.030} \log P}$ & MW \\
        \cite{di_benedetto_pulsational_1994} & BW, SBCR & $\mathrm{1.139_{\pm0.010} + 0.716_{\pm0.010} \log P}$ & LMC \\
        \cite{laney_radii_1995} & BW, SBCR indep. & $\mathrm{1.821_{\pm0.008} + 0.751_{\pm0.026}(\log P - 1)}$ & MW \\
        \cite{gieren_calibrating_1999} & BW, SBCR & $\mathrm{1.146_{\pm0.025} + 0.680_{\pm0.017} \log P}$ & MW, LMC, SMC \\
        \cite{turner_distance_2002} & BW, SBCR indep. & $\mathrm{1.071_{\pm0.025} + 0.747_{\pm0.028} \log P}$ & MW \\
        \cite{kervella_2004_pr_relation} & BW, SBCR indep. & $\mathrm{1.091_{\pm0.011} + 0.767_{\pm0.009} \log P}$ & MW \\
        \cite{groenewegen_projection_2007} & BW, SBCR indep. & $\mathrm{1.134_{\pm0.034} + 0.686_{\pm0.036} \log P}$ & MW \\
        \cite{groenewegen_pr_2013} & BW, SBCR & $\mathrm{1.136_{\pm0.014} + 0.651_{\pm0.012} \log P}$ & MW, LMC, SMC \\
        \cite{gallenne_observational_2017} & BW, SBCR & $\mathrm{1.489_{\pm0.002} + 0.684_{\pm0.007}(\log P-0.517)}$ & MW, LMC, SMC \\
        \cite{trahin_inspecting_2021} & BW, SBCR & $\mathrm{1.763_{\pm0.003} + 0.653_{\pm0.012}(\log P-0.9)}$ & MW \\
        This work & BW, SBCR indep. & $\mathrm{1.133_{\pm 0.044} + 0.688_{\pm 0.038} logP}$ & MW \\
        \hline
        \hline
        \cite{caccin_improvement_1981} & CORS, wo. $\Delta B$ & $\mathrm{1.177_{\pm0.058} + 0.654_{\pm0.052} \log P}$ & MW \\
        \cite{sollazzo_Cepheid_1981} & CORS, w $\Delta B$ & $\mathrm{1.167_{\pm0.024} + 0.700_{\pm 0.020} \log P}$ & MW \\
        \cite{ripepi_Cepheid_1996} & CORS, w. $\Delta B$ & $\mathrm{1.263_{\pm0.033} + 0.606_{\pm0.037} \log P}$ & MW \\
        \cite{ripepi_Cepheid_1996} & CORS, wo. $\Delta B$ & $\mathrm{1.226_{\pm0.026} + 0.622_{\pm0.029} \log P}$ & MW \\
        \cite{ruoppo_improvement_2004} & CORS, wo. $\Delta B$ & $\mathrm{1.18_{\pm0.08} + 0.69_{\pm0.09} \log P}$ & MW \\
        \cite{ruoppo_improvement_2004} & CORS, w $\Delta B$ & $\mathrm{1.19_{\pm0.09} + 0.74_{\pm0.11} \log P}$ & MW \\
        \cite{molinaro_new_nodate} & CORS, w $\Delta B$ & $\mathrm{1.10_{\pm0.03} + 0.75_{\pm0.02} \log P}$ & MW \\
        \hline
        \hline
        \cite{karp_hydrodynamic_1975} & Th & $\mathrm{1.07 + 0.72 \log P}$ & MW \\
        \cite{cogan_radii_1978} & Th & $\mathrm{1.17 + 0.70 \log P}$ & MW \\
        \cite{fernie_survey_1984} & Th & $\mathrm{1.179_{\pm0.006} + 0.692_{\pm0.006} \log P}$ & MW \\
        \cite{bono_theoretical_1998} & Th, canonical & $\mathrm{1.188_{\pm0.008} + 0.655_{\pm0.006} \log P}$ & MW \\
        \cite{bono_theoretical_1998} & Th, non canonical & $\mathrm{1.174_{\pm0.009} + 0.647_{\pm0.006} \log P}$ & MW \\
        \cite{bono_theoretical_1998} & Th, canonical & $\mathrm{1.192_{\pm0.009} + 0.666_{\pm0.007} \log P}$ & LMC \\
        \cite{bono_theoretical_1998} & Th, non canonical & $\mathrm{1.183_{\pm0.009} + 0.653_{\pm0.006} \log P}$ & LMC \\
        \cite{bono_theoretical_1998} & Th, canonical & $\mathrm{1.199_{\pm0.010} + 0.670_{\pm0.008} \log P}$ & SMC \\
        \cite{bono_theoretical_1998} & Th, non canonical & $\mathrm{1.183_{\pm0.009} + 0.661_{\pm0.006} \log P}$ & SMC \\
        \cite{alibert_period_1999} & Th & $\mathrm{1.143 + 0.714 \log P}$ & MW \\
        \cite{alibert_period_1999} & Th & $\mathrm{1.142 + 0.717 \log P}$ & LMC \\
        \cite{alibert_period_1999} & Th & $\mathrm{1.129 + 0.709 \log P}$ & SMC \\
        \cite{petroni_classical_2003} & Th & $\mathrm{1.191_{\pm0.006} + 0.654_{\pm0.005} \log P}$ & MW \\
        \cite{deSomma_2020} & Th, canonical & $\mathrm{1.142_{\pm0.004} + 0.702_{\pm0.003} \log P}$ & MW \\
        \cite{deSomma_2020} & Th, non canonical & $\mathrm{1.128_{\pm0.005} + 0.685_{\pm0.003} \log P}$ & MW \\
        \cite{deSomma_2022} & Th, canonical & $\mathrm{1.153_{\pm0.006} + 0.703_{\pm0.005} \log P}$ & LMC \\
        \cite{deSomma_2022} & Th, non canonical & $\mathrm{1.137_{\pm0.006} + 0.690_{\pm0.004} \log P}$ & LMC \\
        \cite{deSomma_2022} & Th, canonical & $\mathrm{1.146_{\pm0.007} + 0.703_{\pm0.005} \log P}$ & SMC \\
        \cite{deSomma_2022} & Th, non canonical & $\mathrm{1.131_{\pm0.007} + 0.690_{\pm0.005} \log P}$ & SMC \\
        \hline
    \end{tabular}
    \tablefoot{The expressions are in terms of $\mathrm{\log(R/R_{\odot})}$. For some of the relations, the errors on the coefficients are not indicated in the original paper. All PR relations listed here were derived for fundamental classical Cepheids.}
    \label{tab:PR_relations}
\end{table*} 

\subsection{The BW method}
\label{subsec: The BW method+pr}
The Baade-Wesselink (BW) method was first proposed by Lindemann \citep{lindemann_1918} based on the idea of verifying the pulsation hypothesis of variable stars. It was later put into practice by van Hoof \citep{van_Hoof_1945}, who first applied it to calculate the distance of three Cepheids, and later by Wesselink \citep {wesselink_1969}. The principle is as follows: The distance ($\mathrm{d}$) of a pulsating star can be deduced by measuring the limb-darkened angular diameters ($\mathrm{\theta_{LD}}$) over the entire pulsation cycle and comparing them to the variation in stellar radius ($\mathrm{\Delta R}$) deduced from the integration of the radial velocity and using a projection factor \citep{nardetto_2004}. We then have 
\begin{equation}
    \mathrm{\theta_{LD} = \bar{\theta}_{LD} + \frac{\Delta R}{d}}
    \label{eq:bw_method}.
\end{equation}
By fitting this relation, we can determine the distance and mean dimension of the star ($\mathrm{\bar{\theta}_{LD}}$ and $\mathrm{\bar{R}}$). There are three different versions of the BW method, depending on how the variation in the angular diameter is determined: first, directly by interferometry \citep{kervella_2004_I_distances}, second, using an SBCR \citep{storm_2011_A94_I}, and third, using the method called spectro-photo-interferometry for pulsating star (SPIPS)  \citep{merand_2015_spips}, which combines several photometric bands, velocimetry, interferometry, and surface brightness derived using ATLAS9 atmospheric models \citep{castelli_2003_atlas9}. In this version, the authors computed a global fit of all the Cepheid parameters to derive its physical parameters \citep{gallenne_observational_2017, trahin_inspecting_2021}. \cite{laney_radii_1995} used a slightly different method that was still based on the BW concepts. Introduced by \cite{balona_1977}, this method solely relies on the magnitude-colour diagram in the infrared domain, for which the light curves are entirely dominated by variations in the stellar radius. They derived the average radius using a maximum likelihood method. In the first part of Tab.~\ref{tab:PR_relations}, we list the PR relation found in the literature for which the BW method was used to derive the mean stellar radius of the Cepheids. We considered \cite{fernie_survey_1984}, who compiled PR relations, PR relations that used an SBCR \citep{moffett_observational_1987, rojo_arellano_new_1994, di_benedetto_pulsational_1994, gieren_calibrating_1999, groenewegen_pr_2013, gallenne_observational_2017}, PR relations for which $\mathrm{\theta_{LD}}$ was derived using direct interferometry \citep{kervella_2004_pr_relation, groenewegen_projection_2007}, and different versions of the BW method \citep{laney_radii_1995, turner_distance_2002, trahin_inspecting_2021}. \\

\begin{figure}[h]
    \centering
    \includegraphics[width=1.0\linewidth]{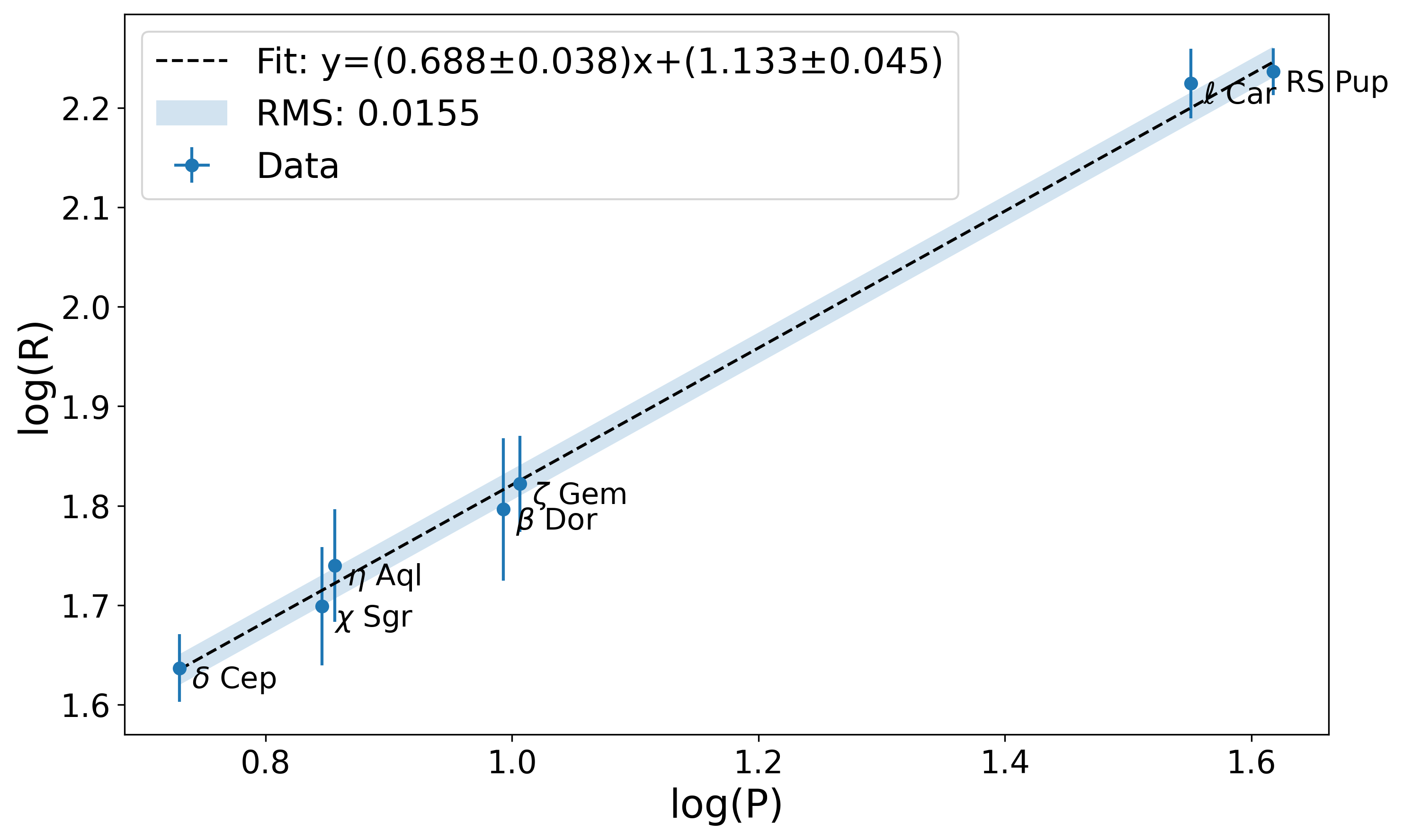}
    \caption{Period-radius relation using data of seven Cepheids based on the interferometric angular diameter curve of \cite{bailleul_2025}.}
    \label{fig:pr_relation}
\end{figure}

Following \cite{bailleul_2025}, we also derived a new period-radius relation for MW Cepheids and derived the mean value of the angular diameter directly from the observed interferometric angular diameter curves, combined with a distance based on various estimates in the literature (listed in Table~\ref{tab:distance_cep}). 
\begin{table}[]
    \caption{Distances in parsec of seven Cepheids from the literature.}
    \centering
    \renewcommand{\arraystretch}{1.2}
    \begin{tabular}{lll}
        \hline
        \multirow{5}{*}{RS Pup}         & \textit{1756 $\pm$ 53} & (1) \\
                                        & 1704.42$^{50.84}_{59.19}$ & (13) \\
                                        & 1695 $\pm$ 62 & (2) \\
                                        & 1910 $\pm$ 80 & (3) \\
                                        & 1641 $\pm$ 148 & (7) \\ \hline 
                                        
        \multirow{6}{*}{$\ell$ Car}     & \textit{503 $\pm$ 27} & (1) \\
                                        & 514.97$^{28.99}_{547.91}$ & (13) \\
                                        & 533 $\pm$ 25 & (2) \\
                                        & 463$^{123}_{83}$ & (4) \\ 
                                        & 603$^{24}_{19}$ & (5) \\
                                        & 566$^{24}_{19}$ & (6) \\ \hline
                                        
        \multirow{8}{*}{$\zeta$ Gem}    & \textit{325 $\pm$ 23} & (1) \\
                                        & 374 $\pm$ 13 & (2) \\
                                        & 370 $\pm$ 33 & (7) \\
                                        & 358$^{146}_{80}$ & (4) \\
                                        & 422$^{61}_{47}$ & (8) \\
                                        & 369$^{24}_{21}$ & (9) \\
                                        & 360$^{25}_{22}$ & (10) \\
                                        & 362 $\pm$ 38 & (11) \\ \hline
                                        
        \multirow{5}{*}{$\beta$ Dor}    & \textit{341 $\pm$ 16} & (1) \\
                                        & 349 $\pm$ 15 & (2) \\
                                        & 316 $\pm$ 28 & (7) \\
                                        & 345$^{175}_{80}$ & (5) \\
                                        & 318$^{74}_{50}$ & (4) \\ \hline
                                        
        \multirow{7}{*}{$\eta$ Aql}     & \textit{272 $\pm$ 14} & (1) \\
                                        & 272.17$^{11.31}_{14.38}$ & (13) \\
                                        & 272 $\pm$ 11 & (2) \\
                                        & 256 $\pm$ 23 & (7) \\
                                        & 360$^{175}_{89}$ & (4) \\
                                        & 276$^{55}_{33}$ & (5) \\
                                        & 320 $\pm$ 32 & (11) \\ \hline
                                        
        \multirow{3}{*}{X Sgr}          & \textit{356 $\pm$ 17} & (1) \\
                                        & 354.35$^{15.80}_{20.95}$ & (13) \\
                                        & 350 $\pm$ 13 & (2) \\
                                        & 330$^{148}_{78}$ & (4) \\ \hline
                                        
        \multirow{5}{*}{$\delta$ Cep}   & \textit{281 $\pm$ 11} & (1) \\
                                        & 281.19$^{9.72}_{10.35}$ & (13) \\
                                        & 266 $\pm$ 12 & (2) \\
                                        & 244 $\pm$ 22 & (7) \\
                                        & 301$^{64}_{45}$ & (4) \\
                                        & 273$^{12}_{11}$ & (12) \\ \hline
    \end{tabular}
    \tablebib{
    (1)~\cite{gaia_EDR3_2020}; (2)~\cite{Skowron_2019}; (3)~\cite{kervella_2024_rsPup}; (4)~\cite{perryman_1997_hypparcos}; (5)~\cite{kervella_2004_I_distances}; (6)~\cite{kervella_2004_lCar_paper}; (7)~\cite{luck_2011}, (8)~\cite{Leeuwen_2007_newHipparcosPllx}; (9)~\cite{Leeuwen_2007_cepDistances}; (10)~\cite{benedict_2007}; (11)~\cite{lane_2002}; (12)~\cite{benedict_2002}, (13)~\cite{bailer_jones_2023}.
    }
    \tablefoot{Following \cite{luck_2011} , we took uncertainties of 9\% for their distance measurements.
    The distances are indicated in parsec. As no zero-point correction is available for all the seven stars, we decided to not use the \textit{Gaia} parallax (distances in italics), but the geometric \cite{bailer_jones_2023} distances when available.}
    \label{tab:distance_cep}
\end{table}
For the seven Cepheids presented by \cite{bailleul_2025}, we first fitted the limb-darkened angular diameter curves using a second-order polynomial fit, and we estimated the error associated with the mean value using a bootstrapping method. The errors associated with the mean angular diameter, the dispersion in distance, and the errors associated with each distance were taken into account in the calculation of the error on the mean radius. The logarithm of the radius is plotted as a function of the period derived from \cite{trahin_inspecting_2021} in Fig.~\ref{fig:pr_relation}. To fit the relation, we used a least-squares fitting method that considered uncertainties on the y-axis and the barycenter of the data points. In this method, the result initially had the following formalism: $\mathrm{y=\alpha}(x-\bar{x})+\mathrm{\beta}$. We then converted the relation into 
$\mathrm{y=}ax+b$ using $a=\mathrm{\alpha}$ and $b=\mathrm{-\alpha}\bar{x} +\mathrm{\beta}$, and we derived the uncertainties from the covariance matrix ($C$) provided by the fitting routine as $\sigma_a =\sqrt{C_{00}}$, $\sigma_b =\sqrt{\bar{x}^2 C_{00}+ C_{11} - 2 \bar{x} C_{01}}$. We obtained the following relation (see also Fig.~\ref{fig:pr_relation}):
\begin{equation}
    \mathrm{\log(R/R_{\odot}) = 0.688_{\pm 0.038} log(P) + 1.133_{\pm 0.044}}
    \label{eq:pr}.
\end{equation}
with a root mean square (RMS) of 0.015. This relation agrees very well with previous empirical relations from the BW method \citep{gieren_calibrating_1999, groenewegen_pr_2013, gallenne_observational_2017} and with theoretical predictions for MW stars \citep{fernie_survey_1984, bono_theoretical_1998,alibert_period_1999, petroni_classical_2003}.

\subsection{The CORS method}
\label{subsec:The CORS method}
The CORS method, named after its authors (Caccin, Onnembo, Russo and Sollazo; \cite{caccin_improvement_1981}), is based on the same hypothesis as the BW method, with the difference that it allows us to determine the radius (instead of the radius variation in BW) of the Cepheids over the pulsation cycle. The method introduces a term $\Delta B$, however, that is directly related to the surface brightness of Cepheids throughout the cycle. In the original CORS method, the $\Delta B$ term was neglected. Later studies \citep{sollazzo_Cepheid_1981, ripepi_1997_CORS} showed that the radius can be determined better when this term is taken into account. 
The method was then modified by including the $\Delta B$ term using either the SBCR derived by \cite{barnes_1976} \citep{ripepi_1997_CORS} or using theoretical grids of effective temperature \citep{ruoppo_improvement_2004, molinaro_new_nodate}.
All the PR relations associated with this method are listed in the second part of Table~\ref{tab:PR_relations}.

\subsection{Theoretical method}
\label{subsec:Theoretical method}
By combining stellar models with a theory of stellar pulsation, one can derive a theoretical radius for Cepheids that takes the stellar parameters of the stars into account, such as the effective temperature, the mass, or the luminosity class. Many authors used this \citep{karp_hydrodynamic_1975,cogan_radii_1978,fernie_survey_1984,petroni_classical_2003, deSomma_2022} with different stellar pulsation and evolution codes. Taking into account the metallicity, \cite{bono_theoretical_1998} and \cite{alibert_period_1999} also derived theoretical PR relations for LMC and SMC stars. All these PR relations are listed in the third part of Table.~\ref{tab:PR_relations}.
\newline

In addition, we note that \cite{pilecki_2018} derived completely independent radius measurements from double-lined eclipsing binaries in the LMC. Their method is very promising and provided the radii of three fundamental-mode Cepheids, which is not sufficient to fit the data and obtain an accurate relation on our side. Furthermore, the authors did not provide a PR relation.\\

\begin{figure}[h!]
    \centering
    \includegraphics[width=1\linewidth]{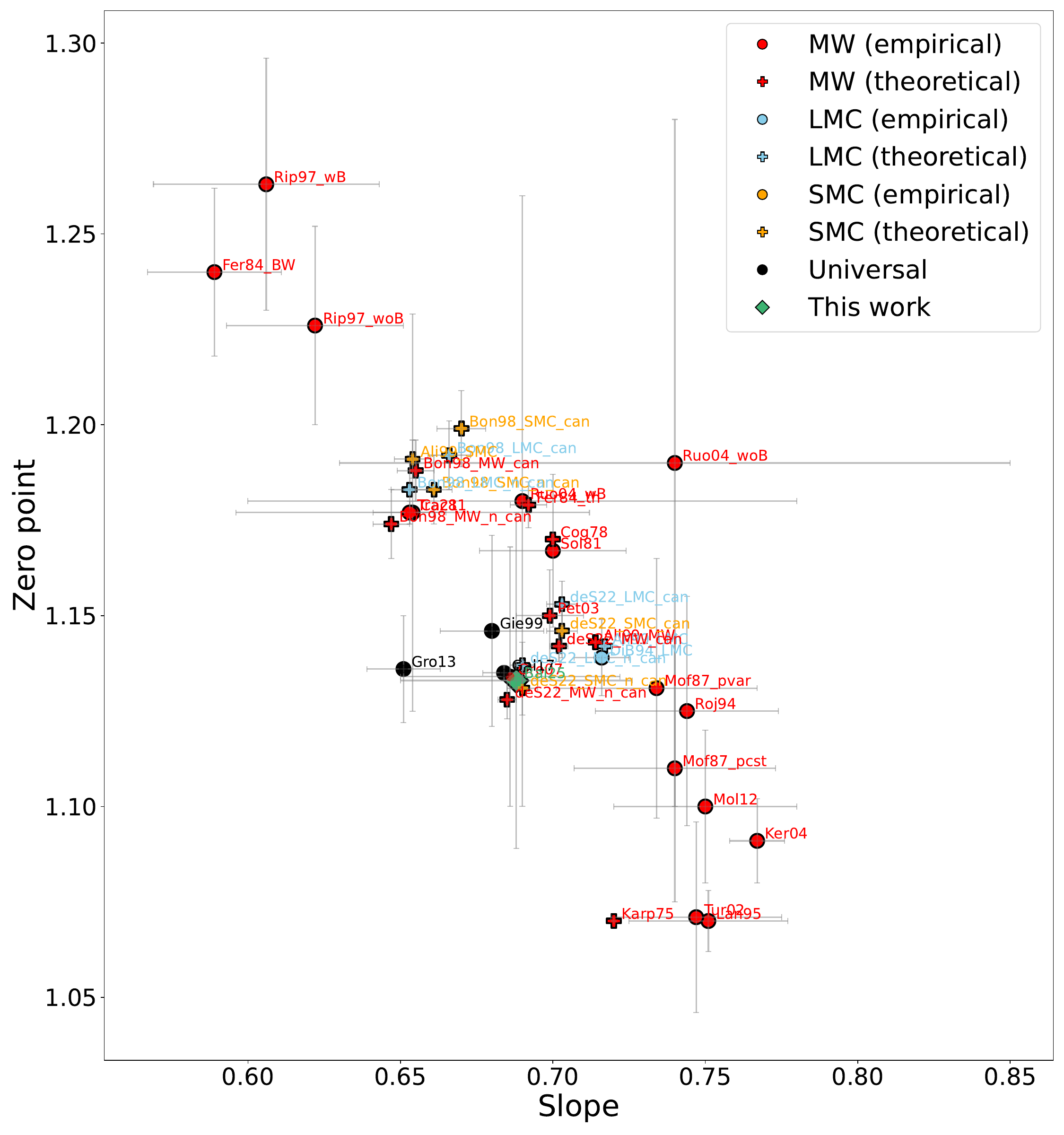}
    \caption{Slope and zero points of the different period-radius relations listed in Tab.~\ref{tab:PR_relations}.}
    \label{fig:ZP_vs_Slope_PRrelations}
\end{figure}

In Fig.~\ref{fig:ZP_vs_Slope_PRrelations} we compare the slope and the zero point of all the PR relations listed in Tab.~\ref{tab:PR_relations}. The blue points show PR relations for LMC Cepheids, the orange ones for SMC Cepheids while the red points are all the PR relations for MW Cepheids. The three black dots correspond to PR relations that are valid for the three galaxies. The green diamond is the PR relation derived in this work.
The slope and zero point are correlated, which is due to the fitting method used in these studies, which does not take the barycenter of the measurements into account. Importantly, we find no evidence that these relations depend on metallicity.

There is also no evidence in the literature that the PR relation depends on metallicity. \cite{gallenne_observational_2017} derived a period-radius (PR) relation for MW, LMC, and SMC Cepheids and found no effect of the metallicity on the relation. This was also an important conclusion of \cite{gieren_calibrating_1999} based on the SBCR approach.

\section{Distances}
\label{sec:distances}
\begin{figure}[h!]
    \centering
    \includegraphics[width=1.15\linewidth]{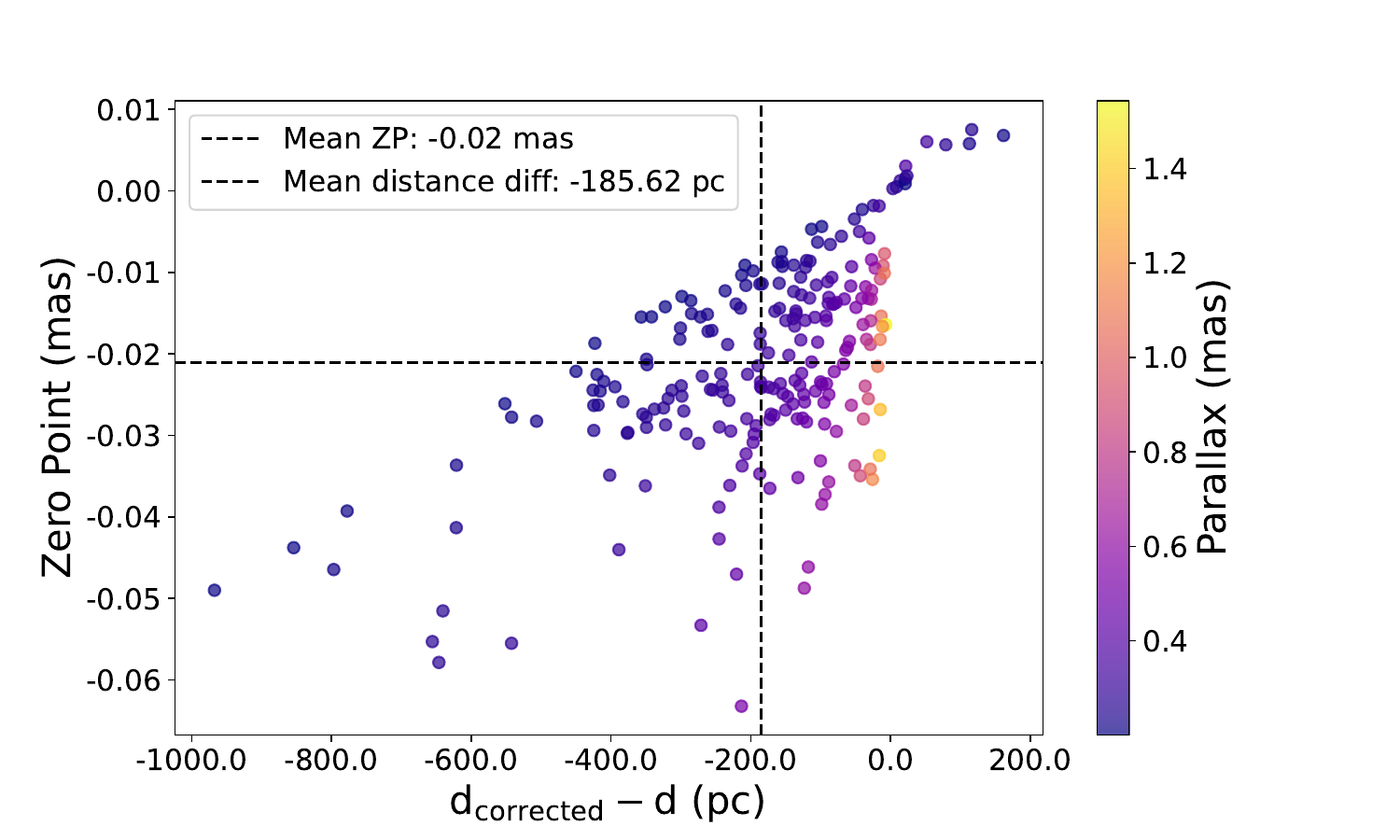}
    \caption{Gaia zero-point offset as a function of the distance difference that it implies. The colour represents the parallax. Only the 228 selected Cepheids (after the whole selection process shown in Fig.~\ref{fig:data_selection}) are shown here.}
    \label{fig:d_correction_impact}
\end{figure}

For the Cepheids in the Milky Way, we adopted the parallaxes from \textit{Gaia} EDR3 \citep{gaia_EDR3_2020} and inverted them to obtain the Cepheid distances. As explained in Sect.~\ref{sec:data_selection}, since we restricted the sample to stars with parallax uncertainties below 10\%, inverting the parallax to estimate the distance is approximately unbiased over this range \citep{bailer_2021}. A correction was applied to all parallaxes according to \cite{lindegren_2021one}, however. We used a dedicated Python code\footnote{\url{https://gitlab.com/icc-ub/public/gaiadr3_zeropoint}} to derive all the zero-point offsets for the different MW Cepheids in our sample. As described by \cite{lindegren_2021one}, interpolations were only calibrated in the following ranges: 6 < \texttt{phot\_g\_mean\_mag} < 21, 1.1 < \texttt{nu\_eff\_used\_in\_astrometry} < 1.9 and 1.24 < \texttt{pseudocolour} < 1.72. Only two Cepheids fell outside these intervals and were therefore rejected from the sample. We found an average parallax offset of -0.02 mas, which corresponds to a distance closer to the Cepheids of {185} pc on average (see Fig.~\ref{fig:d_correction_impact}). This value agrees with the global parallax bias measured from quasars (-17$\mathrm{\mu}$as) mentioned by \cite{gaia_collab_vallenari_2023_instru}.
We also compared our distance determination to the distances of \cite{bailer_jones_2023} and found a negligible median difference of 13.98 parsecs.

\begin{figure}[h!]
    \centering
    \includegraphics[width=1\linewidth]{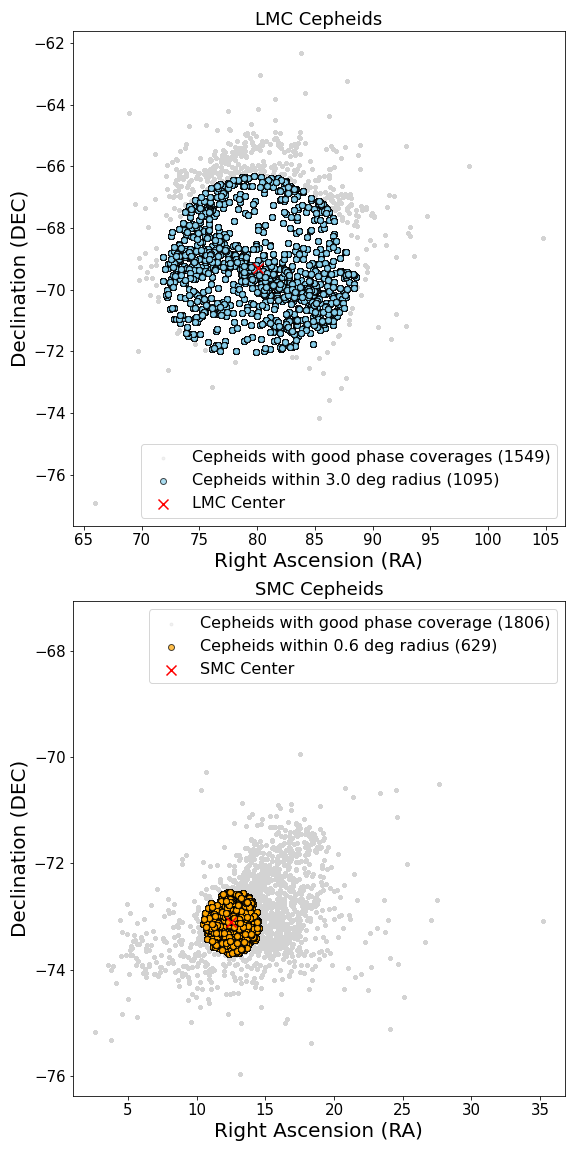}
    \caption{LMC and SMC Cepheids within 3 and 0.6 degrees, respectively, from the clouds centre.}
    \label{fig:lmc_smc_radius_selection}
\end{figure}

For LMC and SMC Cepheids, the distances from \textit{Gaia} parallaxes are not accurate enough. We chose to fix the distance to the distances derived from eclipsing binaries by \cite{pietrzynski_2013, pietrzynski_2019} for the LMC and by \cite{graczyk_2020} for the SMC. We selected the stars that were included within a radius around the clouds centre. This was done to mitigate the effect of the 3D geometry of the Magellanic Clouds. We took as coordinates those used in the two papers cited above: $\mathrm{\alpha_{0}(LMC)=80.05^{\circ}}$, $\mathrm{\delta_{0}(LMC)=-69.3^{\circ}}$ and $\mathrm{\alpha_{0}(SMC)=12.54^{\circ}}$, $\mathrm{\delta_{0}(SMC)=-73.11^{\circ}}$. Following the same method as \cite{breuval_2021}, we used a radius of 3° for the LMC and 0.6° for the SMC (see Fig.~\ref{fig:lmc_smc_radius_selection}).
This selection represents a diameter of 1.3 kpc and 5.2 kpc around the respective centres of the SMC and LMC. We then added a correction to the distance of each star, assuming the geometry of the clouds, but also the elongated shape of the SMC \citep{jacyszyn_2016, graczyk_2020}, following the method of \cite{breuval_2021}.

After this selection, 1095 and 628 Cepheids remained for the LMC and SMC, respectively, representing 46\% and 25\% of the DCEP and FUNDAMENTAL Cepheids in the two galaxies.

\section{Surface brightness-colour relation}
\label{sec:surface_brightness_colour_relation}
As highlighted by \cite{wesselink_1969}, the surface brightness $\mathrm{S_{\lambda}}$ (the flux density received per unit of solid angle) of a star can be directly related to its limb-darkened angular diameter $\mathrm{\theta_{LD}}$, and its apparent magnitude corrected for the extinction $\mathrm{m_{\lambda_{0}}}$ with the following formula:
\begin{equation}
\mathrm{
    S_{\lambda} = m_{\lambda_{0}}+5\log(\theta_{LD}).
    }
\end{equation}
 \cite{barnes_1976} later determined a linear relation between the surface brightness (called $\mathrm{F_{\lambda}}$) and the stellar colour index expressed in magnitude. Then, \cite{fouque_1997} derived the following relation: 
\begin{equation}
\mathrm{
    F_{\lambda} = 4.2196-0.1m_{\lambda_{0}}-0.5\log(\theta_{LD}),
    }
    \label{eq:surface_brightness}
\end{equation}
where $4.2196$ is a constant that depends on the solar parameters \citep{mamajek_2015}. 
For the first time, \cite{bailleul_2025} derived an SBCR dedicated to Cepheids using the $\mathrm{G_{BP}}$ and $\mathrm{G_{RP}}$ bands of the \textit{Gaia} DR3 catalogue. We derived inverse SBCRs for MW, LMC and SMC Cepheids for the same combination of colours. 
The mean limb-darkened angular diameter (in milliarcseconds) was derived as follows:
\begin{equation}
    \mathrm{\overline{\theta_{LD}} = 9.298\frac{\overline{R}}{d}},
    \label{eq:theta=R/d}
\end{equation}
where d is the distance in parsec (see Sec.~\ref{sec:distances}), and $\mathrm{\overline{R}}$ is the mean stellar radius in solar radii (from the PR relations; see Sec.~\ref{sec:period_radius_relations}). The constant comes from the conversion between these units. We then obtained the mean surface brightness (e.g., in the $\mathrm{G_{BP}}$ band) using
\begin{equation}
    \mathrm{F_{G_{BP}} = 4.2196-0.1 (m_{G_{BP}})_{0} - 0.5\log_{10}(\overline{\theta_{LD}})}.
    \label{eq:surface_brightness_Gaiamags}
\end{equation}
The colour of a star can be defined simply as the difference between two dereddened apparent magnitudes measured in two different photometric bands. For our purpose, we used the dereddened average magnitude in the $\mathrm{G_{BP}}$ and $\mathrm{G_{RP}}$ bands.
There are different possibilities regarding the way we derive the mean value of the magnitude over the pulsating cycle of the star. In \textit{Gaia}, \texttt{zpmagBP} and \texttt{zpmagRP} are zero points in mag of the final model of the $\mathrm{G_{BP}}$ and $\mathrm{G_{RP}}$ band light curves (LCs). In other words, these values correspond to the zero point of the Fourier adjustments of the LCs. No uncertainties are available for these values, however. \texttt{BPmagavg} and \texttt{RPmagavg} are intensity-averaged magnitudes for which we have a dedicated uncertainty. In the \textit{Gaia} source catalogue, \texttt{phot\_bp\_mean\_mag} and \texttt{phot\_rp\_mean\_mag} are defined as the mean magnitudes in the integrated $\mathrm{G_{BP}}$ and $\mathrm{G_{RP}}$ band, and they are computed from the $\mathrm{G_{BP}}$ and $\mathrm{G_{RP}}$ band mean flux applying the magnitude zero point in the Vega scale. They lack an associated uncertainty. 
We chose to use the values \texttt{zpmagBP} and \texttt{zpmagRP} instead because they are robustly determined from the LCs Fourier fit when the phase coverage is sufficient. To do this, we only considered stars with at least one measurement per bin of 0.1 in phase (see Sect.~\ref{sec:data_selection}). We emphasize here that only considering an average magnitude of the data points is not enough because it is biased in some cases by the phase coverage (even when we considered a minimum of one measurement per bin of 0.1 in phase). The uncertainty on the mean value of the magnitude was calculated as the average of the uncertainties associated with each point on the curve, however.

\section{Extinction}
\label{sec:extinction}
\begin{figure}[h!]
    \centering
    \includegraphics[width=1\linewidth]{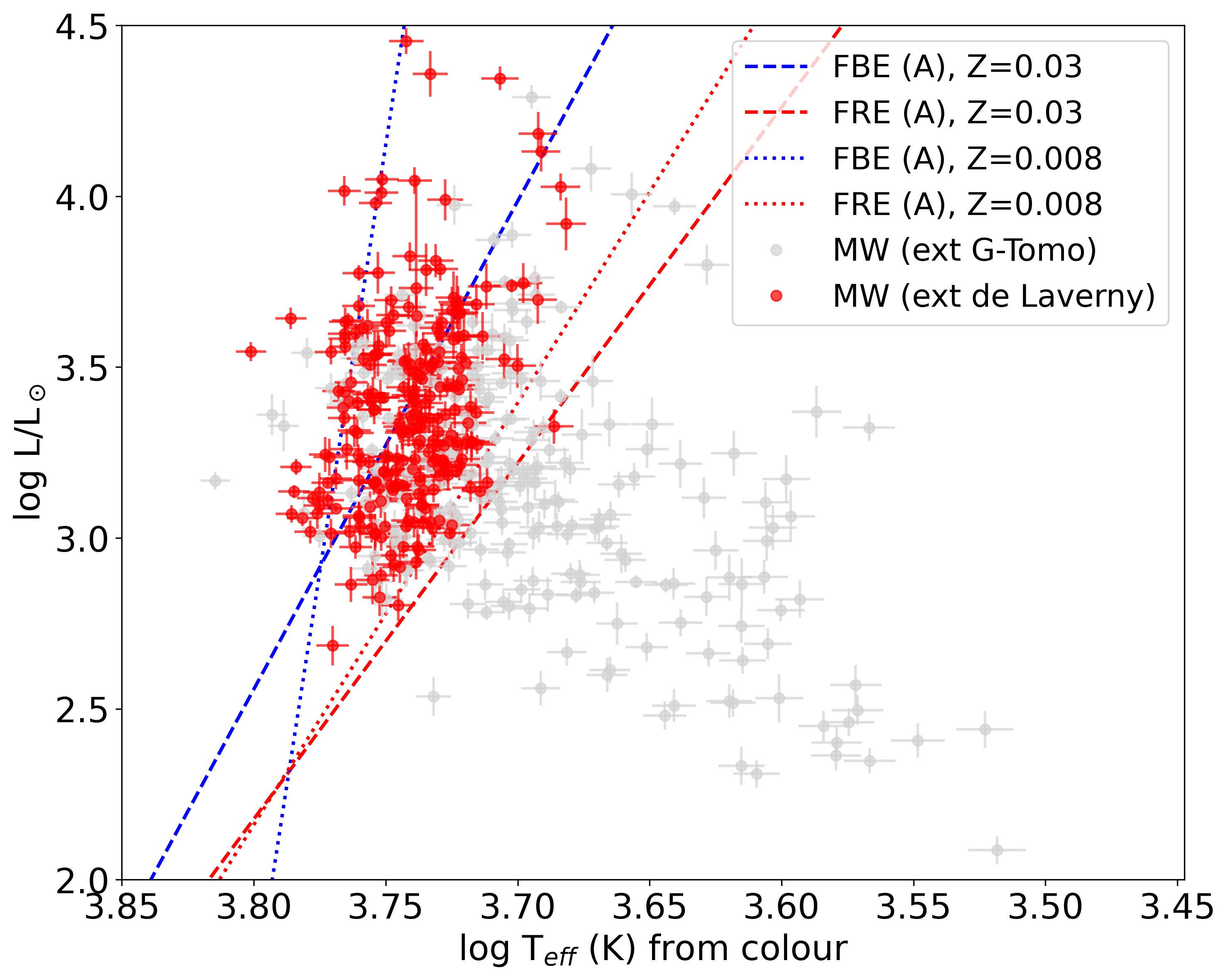}
    \caption{HR diagram for MW Cepheids and position of the red and blue edge of the theoretical IS for Z=0.03 (dashed line) and Z=0.008 (dotted line).}
    \label{fig:IS}
\end{figure}

First, we tried to use the code G-Tomo \citep{lallement_2022} that is available on GitHub\footnote{\url{https://github.com/explore-platform/g-tomo}} to derive the extinction of all the stars, considering the coordinate of the star and its distance as input (following the method described in Sect.~\ref{sec:distances}). More precisely, we decided to use the different extinction maps from \cite{vergely_2022_gtomo_maps} depending on the resolution, the limit in kiloparsec, and the distance of each star. We also rejected all the stars whose distances were larger than 5kpc following the limit of the largest map. We finally found the extinction to be poorly estimated for a large number of Cepheids because their position did not follow the instability strip (IS). 
We chose to use the extinction values ($\mathrm{E(G_{BP}-G_{RP})}$) from de Laverny et al. (in preparation, private communication), derived from the GSPspec/DR3 stellar atmospheric parameters \citep{recioBlanco_2023}. These extinctions were estimated by comparing the observed \textit{Gaia} DR3 ($\mathrm{G_{BP}-G_{RP}}$) colour with theoretical colours computed based on \textit{Gaia} spectroscopic data.
We used the following equation to convert $\mathrm{E(G_{BP}-G_{RP})}$ into $\mathrm{E(B-V)}$: $\mathrm{E(B-V)=E(G_{BP}-G_{RP})/1.392}$ \citep{riello_2021}. The comparison between G-Tomo and de Laverny is given in Fig.~\ref{fig:IS} and confirms the better estimates by this second work.
We plot the theoretical position of the red and blue edge of the IS found in \cite{deSomma_2022} for Z=0.008 and Z=0.03 in a Hertzsprung-Russell (HR) diagram. We used Tab.~1 of \cite{mucciarelli_2021} for giant stars to derive the effective temperature ($\mathrm{T_{eff}}$) from the de-reddened colours. 
 The luminosity was calculated as follows: $\mathrm{\log_{10} \left( \frac{L}{L_\odot} \right) = 0.4 \left( M_{G,\odot} - M_G \right)}$, with $\mathrm{M_G = m_G - 2.70 \times E(B-V) - 5 \log_{10}(d) + 5}$.
To derive the extinction coefficients, we used the following formulas from \cite{ripepi_2019_pl}:
\begin{align}
    &\mathrm{A_{G}} = \mathrm{2.70\,E(B-V)} \\
    &\mathrm{A_{G_{BP}}} = \mathrm{3.50\,E(B-V)} \\
    &\mathrm{A_{G_{RP}}} = \mathrm{2.15\,E(B-V)}.
\end{align} 

\begin{figure}[htbp]
  \centering
  \begin{minipage}[b]{0.49\linewidth}
    \includegraphics[width=\linewidth]{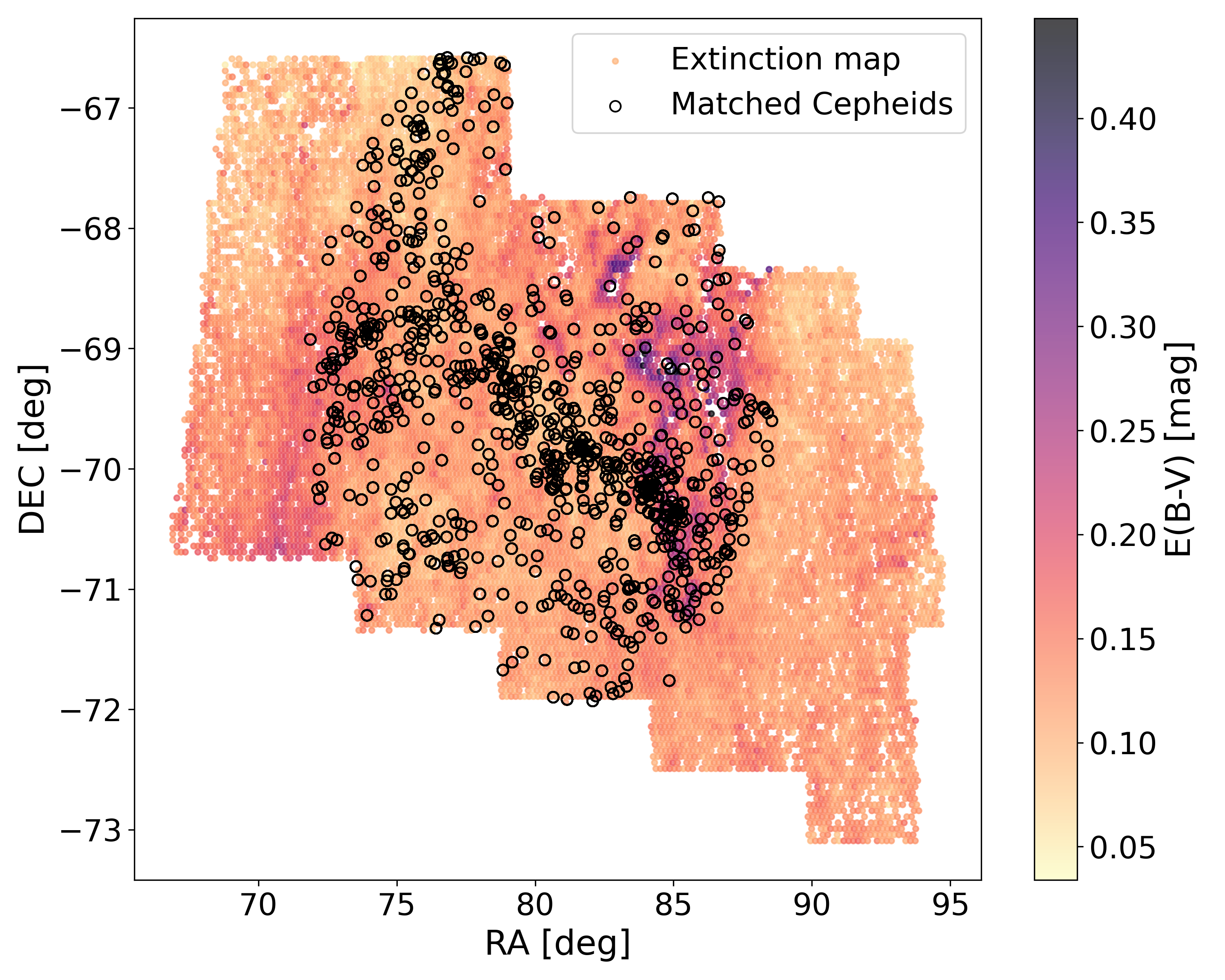}
  \end{minipage}
  \hfill
  \begin{minipage}[b]{0.49\linewidth}
    \includegraphics[width=\linewidth]{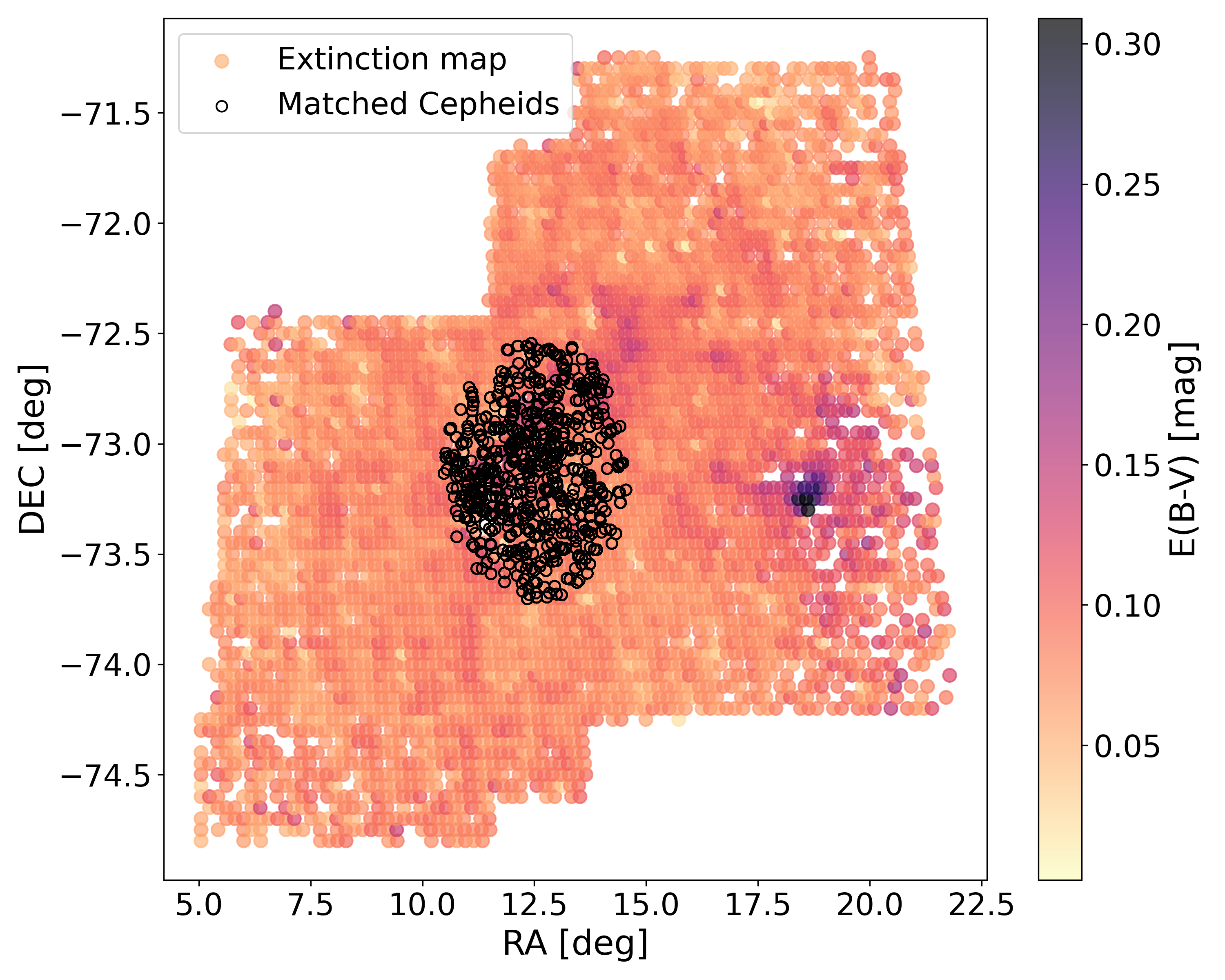}
  \end{minipage}
  \caption{LMC (left) and SMC (right) extinction maps (RA vs. DEC) from \cite{gorski_2020} with matched Cepheids from our sample. The colour map indicates the value of $\mathrm{E(B-V)}$ at a  certain location.}
  \label{fig:ext_map_LMC_SMC}
\end{figure}

For MC Cepheids, we used the result of \cite{wang_2023}, who derived the total-to-selective extinction ratio for red supergiants and classical Cepheids: $\mathrm{R_V(LMC) = 3.40 \pm 0.07}$ and $\mathrm{R_V(SMC) = 2.53 \pm 0.10}$, and the extinction maps from \cite{gorski_2020}, who reported $\mathrm{E(B-V)}$ (see Fig.~\ref{fig:ext_map_LMC_SMC}). 

\section{Metallicity}
\label{sec:metallicity}
We mainly considered three populations of Cepheids: those in the MW, LMC, and SMC. In order to study the metallicity effect on the SBCRs, we estimated the mean Cepheid metallicity in our sample using recent estimates in the literature.

\begin{figure}[h!]
    \centering
    \includegraphics[width=1\linewidth]{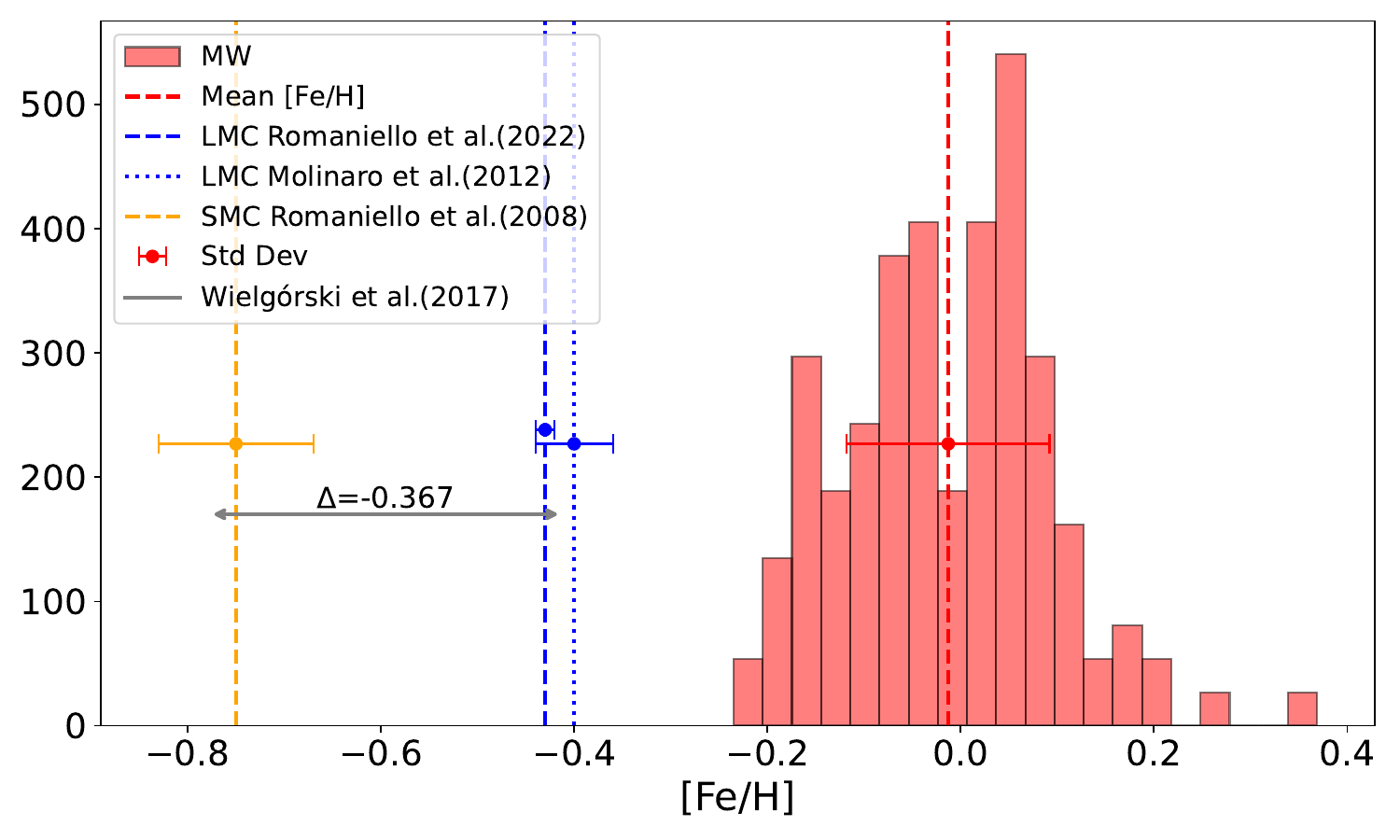}
    \caption{Different metallicities found in the literature for MW Cepheids (red), LMC (blue), and SMC (orange; see Sect.~\ref{sec:metallicity} for details).}
    \label{fig:metallicity_MW_SMC_LMC}
\end{figure}

\cite{hocde_2023} gathered different determinations of the spectroscopic Cepheid metallicity from different sources and re-scaled them to the solar abundance value of A(Fe) = 7.50 \citep{asplund_2009}. This compilation is based on the largest dataset with a homogeneous determination of [Fe/H] of MW Cepheids from \cite{luck_2018} (435 Cepheids), which was used as a reference to correct for the zero point of different samples \citep{genovali_2014, genovali_2015, kovtyukh_2022a, ripepi_2021, trentin_2023} using stars that are in common to derive the mean offset. We cross-matched this compilation and found 131 Cepheids in common with our sample and an average value of [Fe/H]=-0.013 dex ($\sigma=0.105$ dex) for the MW Cepheids (see Fig.~\ref{fig:metallicity_MW_SMC_LMC}). We emphasize that most of our stars (109/131) were homogeneously determined by \cite{luck_2018}.

Few metallicity determinations for LMC and SMC Cepheids are available. \cite{romaniello_2008} found an average value of $-0.33$ dex ($\sigma=0.13$) for the LMC and $-0.75$ dex ($\sigma=0.08$) for the SMC with a sample of 22 and 14 Cepheids, respectively. \cite{romaniello_2022} found a systematic offset from their previous determination of -0.11 dex for the 22 LMC Cepheids they analysed, however. The new study of \cite{romaniello_2022} with 89 LMC Cepheids leads to an average value of $\mathrm{[Fe/H]_{LMC}=-0.43\pm0.01}$, ($\sigma=0.08$).
\cite{molinaro_2012} found an average value of $\mathrm{[Fe/H]_{LMC} = -0.40\pm0.04}$ from three Cepheids in the NGC
1866 cluster. Finally, \cite{wielgorski_2017} calculated an average metallicity difference between the two clouds of about $-0.367\pm0.106$ dex.
We observe a difference in metallicity ([Fe/H]) between the three galaxies. According to the values found in the literature, this difference is approximately 0.35 dex between the SMC and the LMC, and it is 0.402 dex between the LMC and the population of Cepheids in the Milky Way.

The approach presented in this section based on rescaled inhomogeneous large datasets for MW Cepheids and small samples for MC Cepheids will be greatly improved when the metallicity of Cepheids in \textit{Gaia} will be provided by the 4th Data Release.

\section{Results}
\label{sec:results}

\subsection{Period-Wesenheit relations in the Gaia passbands}
\label{sec:Period-Wesenheit relations in the Gaia passbands}
The first aim of this approach is to be as independent as possible of the extinction (especially for the MW Cepheids). We considered that the Cepheids in the three clouds should follow the PW relation. Those that did not follow the relation might be stars with an incorrectly determined magnitude and undetected or incorrectly classified companions (DCEP and FUNDAMENTAL in \textit{Gaia}).
We derived the extinction-free absolute Wesenheit magnitude using the following equation \citep{ripepi_2019_pl}: 
\begin{equation} \mathrm{W_{G}=(G- 5\log_{10}(d) + 5)-1.90\,(G_{BP}-G_{RP})} ,
\end{equation} 
where $\mathrm{G}$ is the apparent magnitude, $\mathrm{d}$ is the distance in parsec, and $\mathrm{(G_{BP}-G_{RP})}$ is the colour index. In this section, all fitting procedure were made taking the barycenter of the measurement and the uncertainties on the y-axis into account (see Sect.~\ref{subsec: The BW method+pr} for details). We arbitrary rejected all stars above 1.75 times the RMS of the relation in order to retain most of the data ($\approx$ 92\%) while removing outliers. We tested clipping thresholds from 2.5$\sigma$ to 1.5$\sigma$, and the resulting slopes and intercepts were all consistent within the uncertainties with our adopted value of 1.75$\sigma$. Fig.~\ref{fig:Wesenheit_vs_log10_period} (top) shows the three PW relations and the rejected stars for each galaxy corresponding to these relations
\begin{align}
&\mathrm{W_{MW}}  = -3.3090_{\pm 0.0242}\,\mathrm{log_{10}(P)} - 2.6595_{\pm 0.0199} \label{eq:PW_MW}\\
&\mathrm{W_{LMC}}  = -3.3185_{\pm 0.0049}\,\mathrm{log_{10}(P)} - 2.4910_{\pm 0.0033} \label{eq:PW_LMC} \\
&\mathrm{W_{SMC}}  = -3.4426_{\pm 0.0062}\,\mathrm{log_{10}(P)} - 2.3168_{\pm 0.0034} \label{eq:PW_SMC} .
\end{align}

Eq.~\ref{eq:PW_MW} is compatible with the result of \cite{ripepi_2019_pl} for fundamental MW Cepheids. We plot in Figure~\ref{fig:Slope_ZeroPoint_vs_Metallicity_PW_SBCR_2x2} (a) the slopes and (b) the zero points of the three relations above as a function of the metallicity values described in Sect.~\ref{sec:metallicity}. The slope appears to be independent of metallicity. In contrast, we observe a linear deviation between the zero points (ZP) and metallicities, which allowed us to adjust the relation between the zero points and average metallicities of the three galaxies (indicated in the plot), 
\begin{equation}
    \mathrm{ZP_{PW}=-0.5289_{\pm 0.0136} [Fe/H] - 2.7153_{\pm 0.0082}}.
    \label{eq:ZP_PW}
\end{equation}

\noindent
\cite{ripepi_2019_pl} showed that the metallicity affects the slope of the PW relation in the \textit{Gaia} bands only little, which agrees with several previous studies (\cite{fiorentino_2013, ngeow_2012, di_Criscienzo_2013, gieren_2018}). More recently, \cite{ripepi_2022} derived a new period-Wesenheit-metallicity relation in the \textit{Gaia} bands and obtained a metallicity coefficient of $c = -0.520_{\pm 0.090}$, which directly affects the zero point of the PW relation. We determined the relation between the zero point and metallicity (see Eq.~\ref{eq:ZP_PW}). By rewriting this equation,  $\mathrm{W=a\log_{10}(P) - 2.7153_{\pm 0.0082} -0.5289_{\pm 0.0136} [Fe/H]}$, with $a$ that depends on the considered galaxies, the metallicity term we derive, $-0.5289_{\pm 0.0136}$, is fully consistent with the value derived by \cite{ripepi_2022} and with the constant term of the relations.

\subsection{Deriving the SBCRs in the MW, LMC, and SMC}
\begin{figure}[]
    \centering
    \includegraphics[width=1\linewidth]{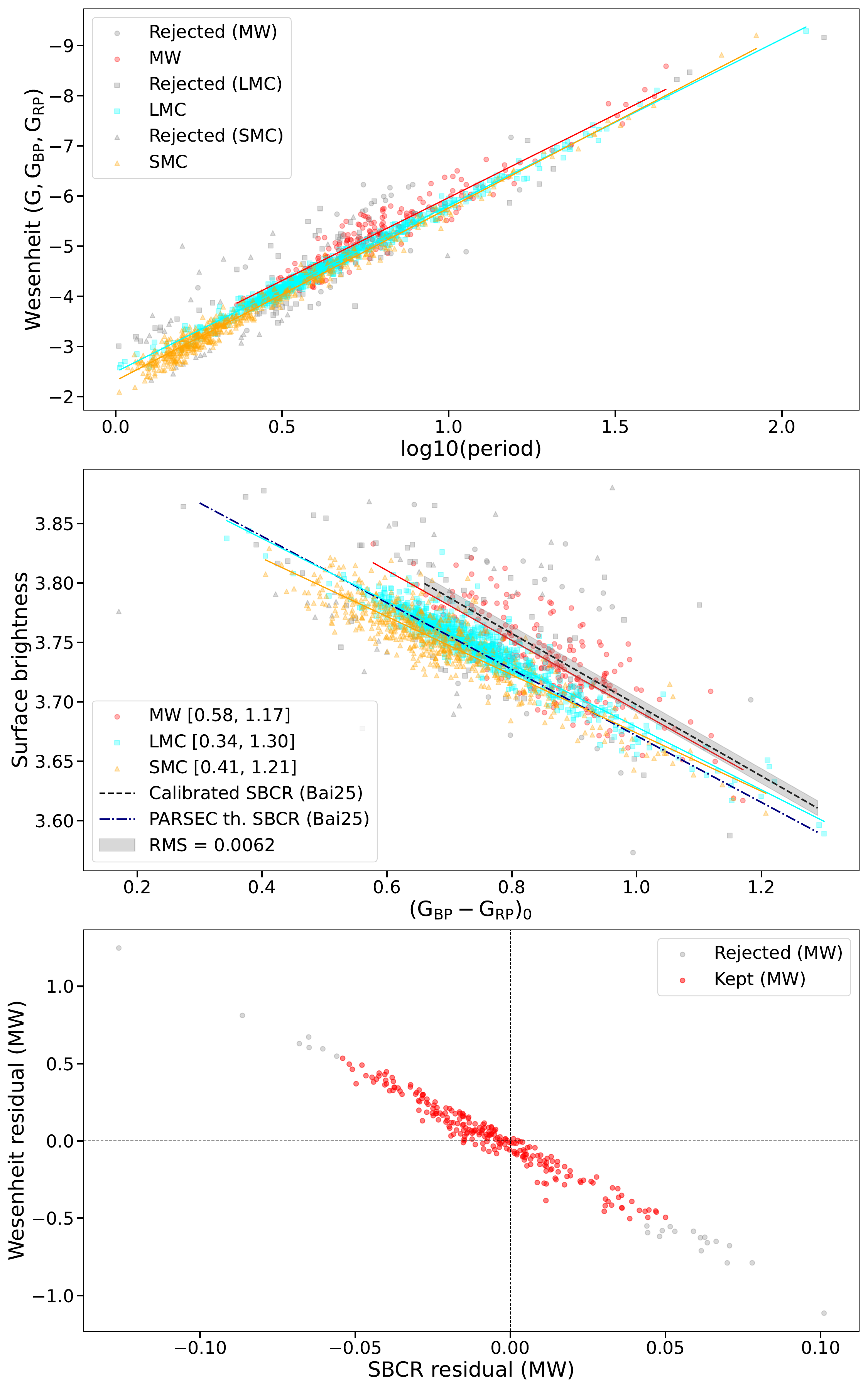}
    \caption{Top: Absolute period-Wesenheit (PW) relations for DCEP and FUNDAMENTAL MW, LMC, and SMC Cepheids. Middle: SBCRs using the Gaia $\mathrm{G_{BP}}$ and $\mathrm{G_{RP}}$ bands. The colour range is indicated in the legend. Bottom: Relation between the PW fit residual and the SBCR residual relative to the calibrated SBCR represented by the dotted black line in the middle panel (only for MW Cepheids).}
    \label{fig:Wesenheit_vs_log10_period}
\end{figure}

After selecting the data based on quality criteria (Sect.~\ref{sec:data_selection}), we determined the distances of each Cepheid (Sect.~\ref{sec:distances}), corrected the magnitudes for the extinction (Sect.~\ref{sec:extinction}), and finally, rejected stars that did not follow the PW relations (Sect.~\ref{sec:Period-Wesenheit relations in the Gaia passbands}). Finally, we derived the average angular diameter of each star using Eq.~\ref{eq:theta=R/d} and the PR relation we derived (Eq.~\ref{eq:pr}), which allowed us to obtain the surface brightness using Eq.~\ref{eq:surface_brightness_Gaiamags} and the colour of each star. Fitting the three relations, we found 
\begin{align}
%\label{eq:sbcr_mw}
&\mathrm{F_{G_{BP},MW}}  = -0.2940_{\pm 0.0175} \mathrm{(G_{BP} - G_{RP})_0} + 3.9870_{\pm 0.0151} \label{eq:sbcr_MW}\\
&\mathrm{F_{G_{BP},LMC}} = -0.2644_{\pm 0.0070} \mathrm{(G_{BP} - G_{RP})_0} + 3.9433_{\pm 0.0053} \label{eq:sbcr_LMC} \\ 
&\mathrm{F_{G_{BP},SMC}} = -0.2450_{\pm 0.0083} \mathrm{(G_{BP} - G_{RP})_0} + 3.9188_{\pm 0.0058} \label{eq:sbcr_SMC} ,
%\label{eq:sbcr_smc}
\end{align}
with an RMS of 0.0223, 0.0096, and 0.0149 respectively. The colour range is indicated in the middle panel of Fig.~\ref{fig:Wesenheit_vs_log10_period}.
\begin{figure}
    \centering
    \includegraphics[width=1\linewidth]{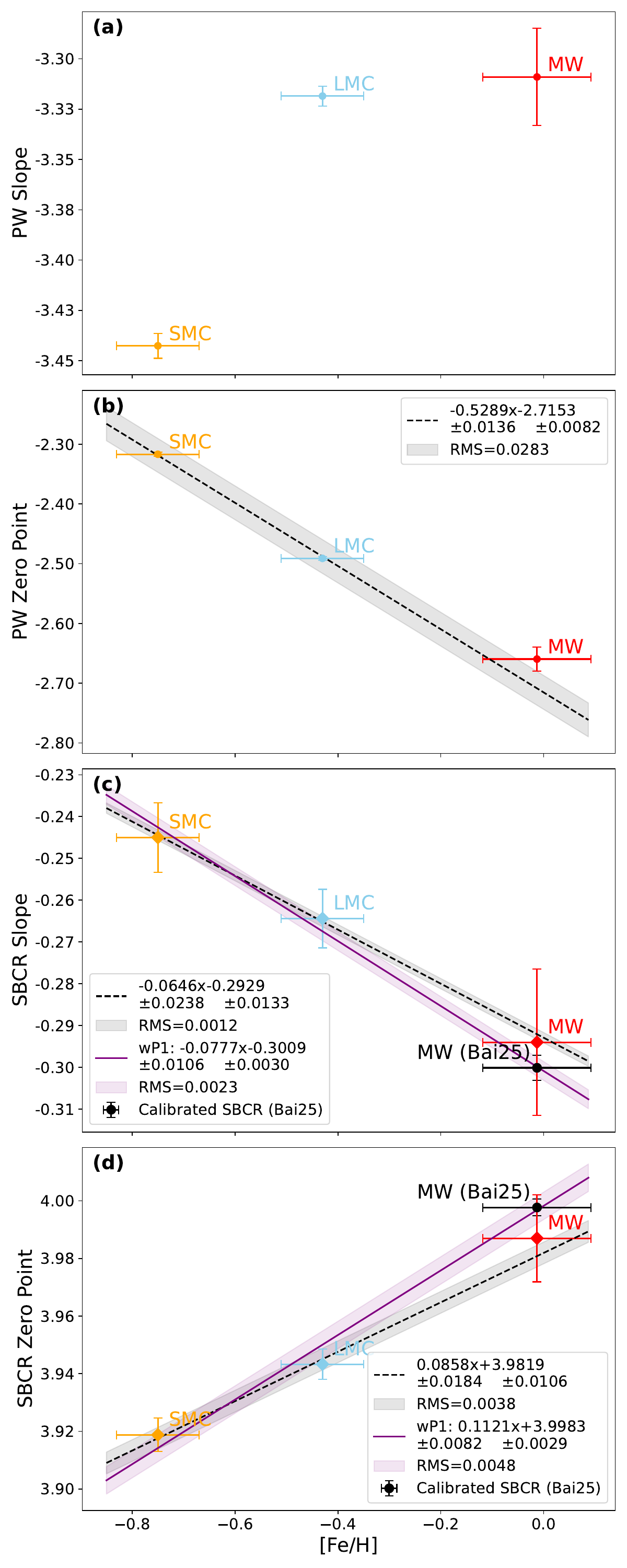}
\caption{Panels (a) and (b): Slopes and zero points from Eqs.~\ref{eq:PW_MW}–\ref{eq:PW_SMC} as a function of [Fe/H]. We adopted the average [Fe/H] value in Sect.~\ref{sec:metallicity} for the MW and the values from \citet{romaniello_2008} and \citet{romaniello_2022} for the SMC and LMC, respectively.  Panels (c) and (d): Same, but for the SBCRs from Eqs.~\ref{eq:sbcr_MW}–\ref{eq:sbcr_SMC}.}
    \label{fig:Slope_ZeroPoint_vs_Metallicity_PW_SBCR_2x2}
\end{figure}
In Fig.~\ref{fig:Wesenheit_vs_log10_period} we overplot the latest SBCR for MW Cepheids calibrated by \cite{bailleul_2025} by interferometry (dashed black line). 
First, the SBCR (MW) derived here agrees well with the previous calibrated one \citep{bailleul_2025} based on direct interferometric measurements. The SBCR (MW) calibrated by interferometry remains the one that should be used in forthcoming studies.
Secondly, the SBCR (LMC) agrees well with the theoretical SBCR for metal-poor Cepheids derived by \cite{bailleul_2025} (see the middle panel of Fig.~\ref{fig:Wesenheit_vs_log10_period}).
Thirdly, we observe that the zero point (as expected) and also the slope of the SBCR are affected by the metallicity. In Figure~\ref{fig:Slope_ZeroPoint_vs_Metallicity_PW_SBCR_2x2} we plot (c) the slopes and (d) the zero points of the three SBCRs as a function of the metallicity. We clearly observe a linear relation between these quantities. We also show the SBCR calibrated by \cite{bailleul_2025} (black dots). In addition, we fitted a linear relation between the slopes, the zero points, and the metallicity using the result of this work (dotted line) and the calibrated SBCR (purple line),
\begin{align}
    & \mathrm{Slope_{SBCR}=-0.0777_{\pm 0.0106} [Fe/H] - 0.3009_{\pm 0.0030}} \\
    & \mathrm{ZP_{SBCR}=0.1121_{\pm 0.0082} [Fe/H] + 3.9983_{\pm 0.0029}}.
\end{align}

The fit was performed without taking the barycenter of the measurement into account.
This result shows that the use of a $\textit{Gaia}$ SBCR, calibrated for stars with solar metallicity, cannot be applied to stars with different metallicities without running the risk of determining the angular diameter incorrectly.
Fourth, we observed and validated the fact that there is a direct link between the residual of the PW(MW) relation and the residual relative to the calibrated SBCR of the SBCR (MW). This indicates that physically, Cepheids that do not correctly follow the PW relation not correctly follow the SBCR either (see Fig.~\ref{fig:Wesenheit_vs_log10_period}, lower panel).
\begin{figure}[h!]
    \centering
    \includegraphics[width=1\linewidth]{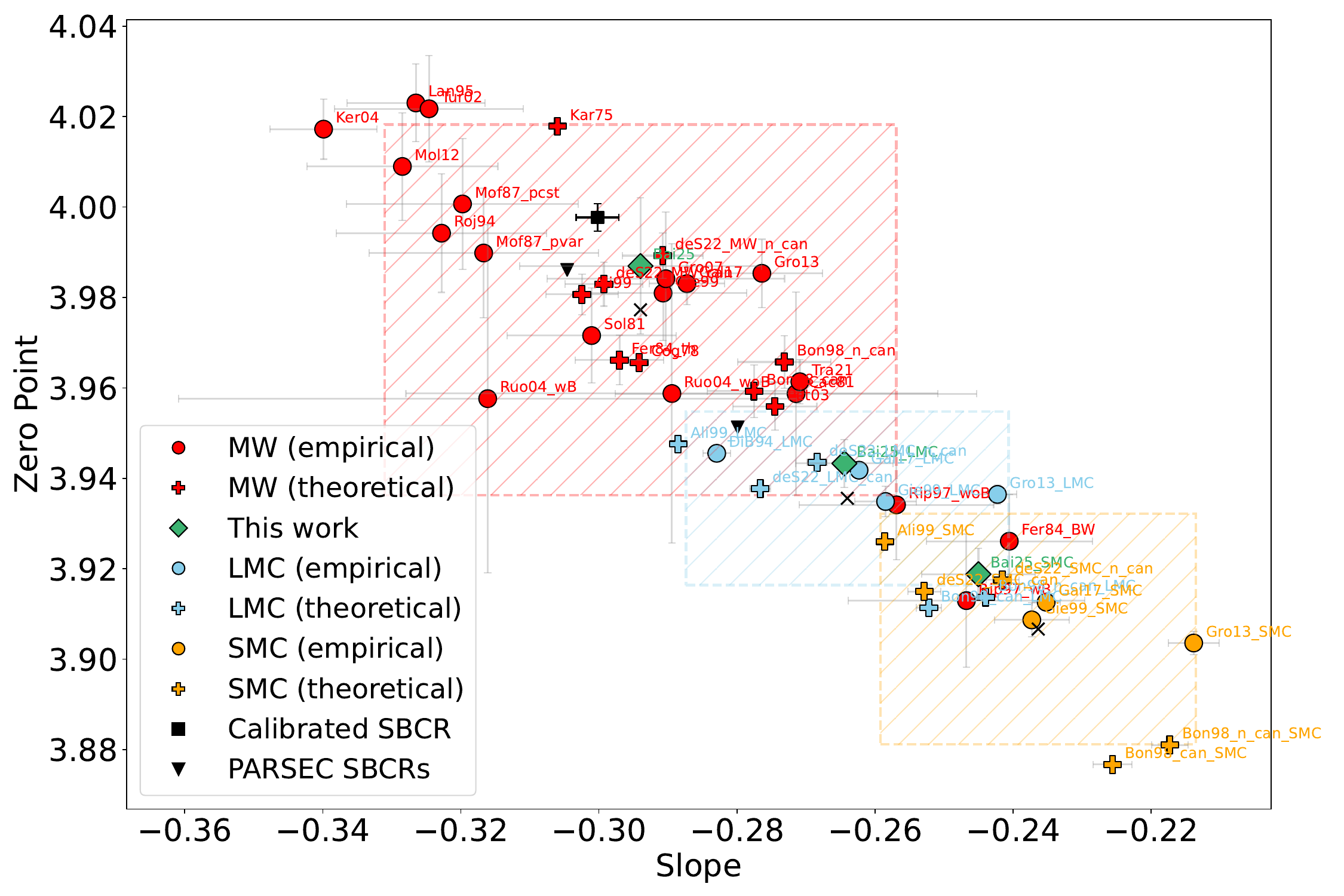}
    \caption{Zero point and slope of the different SBCRs determined using the different PR relations in Tab.\ref{tab:PR_relations}. We show the calibrated and theoretical (PARSEC) SBCRs for solar and metal poor Cepheids determined by \cite{bailleul_2025} (black). The turquoise points indicate the slope and zero point of the SBCRs (MW, LMC, and SMC) using the PR relation calibrated in this paper. In the centre of each dashed rectangle, the mean value is indicated with a black cross. The height (width) of the rectangle is three times the standard deviation ($3 \sigma$) of the zero points found using the various PR relations. Relations without an uncertainty in Table \ref{tab:PR_relations} are shown without uncertainty in this figure.}
    \label{fig:zp_slope_SBCRs}
\end{figure}
Finally, we derived the SBCRs associated with all the PR relation presented in Table \ref{tab:PR_relations}.  The resulting slopes and zero points are plotted in Fig.~\ref{fig:zp_slope_SBCRs}. Again, we find some correlation between the slope and zero point, as in Fig.~\ref{fig:ZP_vs_Slope_PRrelations} for the PR relations.
As described above, however, the trend in Fig.~\ref{fig:ZP_vs_Slope_PRrelations} arises because the different authors did not fit their data taking the barycenters of the measurements into account in x and y, respectively which generates a larger correlation between the fitted slope and the zero point of their relation. Although this trend is observed in PR relations (Fig.~\ref{fig:ZP_vs_Slope_PRrelations}), there is no clear effect of the metallicity, however. Conversely, in Fig.~\ref{fig:zp_slope_SBCRs}, we clearly find three different groups of SBCRs (slope and zero point) for the MW, LMC, and SMC. These results show that the metallicity significantly affects the SBCR, while its effect on PR relations is negligible or even null. Metallic lines are numerous in the large wavelength domain of \textit{Gaia}, which significantly increases the corresponding magnitudes, while the radius derived either from interferometry or SBCR (V, V-K) is known to be weakly affected by the metallicity.

\section{Conclusion}
\cite{bailleul_2025} recently derived an SBCR for the first time based exclusively on \textit{Gaia} photometric bands for fundamental-mode classical Cepheids of solar metallicity. This result is of particular interest because \textit{Gaia} represents one of the largest available Cepheid databases. The authors also demonstrated theoretically that the \textit{Gaia}-based SBCR is highly sensitive to metallicity.
We confirmed these findings empirically for the first time by deriving SBCRs for Cepheids in the Milky Way, Large Magellanic Cloud, and Small Magellanic Cloud. We demonstrated and quantified the effect of metallicity on the slope and zero point of these relations.
With this work, we aim to open a new avenue for the application of the inverse Baade–Wesselink method to Galactic and Magellanic Cloud Cepheids in the \textit{Gaia} sample in order to investigate their physical properties, and in particular, their projection factor \cite{nardetto_2004, nardetto_2007}. The current study is limited by the inhomogeneous and relatively small number of available metallicity measurements for Cepheids. The forthcoming \textit{Gaia} Data Release 4 (DR4) is expected to dramatically improve this situation by providing epoch metallicities for a larger number of stars and with an  improved precision, however.
The overall goal of the Unlockpfactor project \citep{nardetto_2023_pfactor} is to shed light on the projection factor in order to pave the way for the application of the Baade–Wesselink method within the Local Group using ELT spectroscopic instruments, in particular, MOSAIC \citep{mozaic_pello_2024} and ANDES \citep{andes_marconi_2024}.

\begin{acknowledgements}
The authors acknowledge the support of the French Agence Nationale de la Recherche (ANR), under grant ANR-23-CE31-0009-01 (Unlock-pfactor) and the financial support from ``Programme National de Physique Stellaire'' (PNPS) of CNRS/INSU, France.  This research has made use of the SIMBAD and VIZIER (available at \href{http://cdsweb.u-strasbg.fr/}{http://cdsweb.u-strasbg.fr/}) databases at CDS, Strasbourg (France), and of the electronic bibliography maintained by the NASA/ADS system.
This work has made use of data from the European Space Agency (ESA) mission {\it Gaia} (\url{https://www.cosmos.esa.int/gaia}), processed by the {\it Gaia} Data Processing and Analysis Consortium (DPAC, \url{https://www.cosmos.esa.int/web/gaia/dpac/consortium}). Funding for the DPAC has been provided by national institutions, in particular the institutions participating in the {\it Gaia} Multilateral Agreement.
The research leading to these results has received funding from the European Research Council (ERC) under the European Union's Horizon 2020 research and innovation program (projects CepBin, grant agreement 695099, and UniverScale, grant agreement 951549).
This research has used data, tools or materials developed as part of the EXPLORE project that has received funding from the European Union’s Horizon 2020 research and innovation programme under grant agreement No 101004214.
WG gratefully acknowledges financial support for this work from the BASAL Centro de Astrofisica y Tecnologias Afines (CATA) PFB-06/2007, and from the Millenium Institute of Astrophysics (MAS) of the Iniciativa Cientifica Milenio del Ministerio de Economia, Fomento y Turismo de Chile, project IC120009. WG also acknowledges support from the ANID BASAL project ACE210002. Support from the Polish National Science Center grant MAESTRO 2012/06/A/ST9/00269 and 2024/WK/02 grants of the Polish Minstry of Science and Higher Education is also acknowledged. AG acknowledges the support of the Agencia Nacional de Investigaci\'on Cient\'ifica y Desarrollo (ANID) through the FONDECYT Regular grant 1241073.
A.G. acknowledges support from the ANID-ALMA fund No. ASTRO20-0059. B.P. gratefully acknowledges support from the Polish National Science Center grant SONATA BIS 2020/38/E/ST9/00486. This research has been supported by the Polish-French Marie Sk{\l}odowska-Curie and Pierre Curie Science Prize awarded by the Foundation for Polish Science.

\end{acknowledgements}

\renewcommand{\bibname}{References}  % (Optionnel) Pour changer le titre de la biblio
\def\doibase#1{} % Supprime les DOI

\bibliographystyle{aa}
\bibliography{bibtex_P2}

@article{moffett_observational_1987,
	title = {Observational studies of {Cepheids}. {VI} - {Period}-radius relations},
	volume = {323},
	issn = {0004-637X, 1538-4357},
	url = {http://adsabs.harvard.edu/doi/10.1086/165825},
	doi = {10.1086/165825},
	language = {en},
	urldate = {2025-01-21},
	journal = {The Astrophysical Journal},
	author = {Moffett, Thomas J. and Barnes, Iii, Thomas J.},
	month = dec,
	year = {1987},
	pages = {280},
}

@article{molinaro_new_nodate,
	title = {A new determination of the {Period}\{\vphantom{\}}{Radius} relation for {Classical} {Galactic} {Cepheids}.},
	language = {en},
	author = {Molinaro, R and Ripepi, V and Marconi, M and Bono, G and Lub, J and Pedicelli, S and Pel, J W},
    year = {2012},
}

@article{gieren_calibrating_1999,
	title = {Calibrating the {Cepheid} {Period}‐{Radius} {Relation} with {Galactic} and {Magellanic} {Cloud} {Cepheids}},
	volume = {512},
	issn = {0004-637X, 1538-4357},
	url = {https://iopscience.iop.org/article/10.1086/306800},
	doi = {10.1086/306800},
	language = {en},
	number = {2},
	urldate = {2025-01-21},
	journal = {The Astrophysical Journal},
	author = {Gieren, Wolfgang P. and Moffett, Thomas J. and Barnes Iii, Thomas G.},
	month = feb,
	year = {1999},
	pages = {553--557},
}

@article{bono_theoretical_1998,
	title = {On the {Theoretical} {Period}-{Radius} {Relation} of {Classical} {Cepheids}},
	volume = {497},
	issn = {0004637X},
	url = {https://iopscience.iop.org/article/10.1086/311270},
	doi = {10.1086/311270},
	number = {1},
	urldate = {2025-01-21},
	journal = {The Astrophysical Journal},
	author = {Bono, Giuseppe and Caputo, Filippina and Marconi, Marcella},
	month = apr,
	year = {1998},
	pages = {L43--L46},
}

@article{trahin_inspecting_2021,
	title = {Inspecting the {Cepheid} parallax of pulsation using \textit{{Gaia}} {EDR3} parallaxes: {Projection} factor and period-luminosity and period-radius relations},
	volume = {656},
	copyright = {https://creativecommons.org/licenses/by/4.0},
	issn = {0004-6361, 1432-0746},
	shorttitle = {Inspecting the {Cepheid} parallax of pulsation using \textit{{Gaia}} {EDR3} parallaxes},
	url = {https://www.aanda.org/10.1051/0004-6361/202141680},
	doi = {10.1051/0004-6361/202141680},
	urldate = {2025-01-21},
	journal = {Astronomy \& Astrophysics},
	author = {Trahin, B. and Breuval, L. and Kervella, P. and Mérand, A. and Nardetto, N. and Gallenne, A. and Hocdé, V. and Gieren, W.},
	month = dec,
	year = {2021},
	pages = {A102},
}

@article{gallenne_observational_2017,
	title = {Observational calibration of the projection factor of {Cepheids}: {IV}. {Period}-projection factor relation of {Galactic} and {Magellanic} {Cloud} {Cepheids}},
	volume = {608},
	copyright = {https://www.edpsciences.org/en/authors/copyright-and-licensing},
	issn = {0004-6361, 1432-0746},
	shorttitle = {Observational calibration of the projection factor of {Cepheids}},
	url = {http://www.aanda.org/10.1051/0004-6361/201731589},
	doi = {10.1051/0004-6361/201731589},
	urldate = {2025-01-21},
	journal = {Astronomy \& Astrophysics},
	author = {Gallenne, A. and Kervella, P. and Mérand, A. and Pietrzyński, G. and Gieren, W. and Nardetto, N. and Trahin, B.},
	month = dec,
	year = {2017},
	pages = {A18},
}

@article{alibert_period_1999,
	title = {Period - luminosity - color - radius relationships of {Cepheids} as a function of metallicity: evolutionary effects},
	copyright = {Assumed arXiv.org perpetual, non-exclusive license to distribute this article for submissions made before January 2004},
	shorttitle = {Period - luminosity - color - radius relationships of {Cepheids} as a function of metallicity},
	url = {https://arxiv.org/abs/astro-ph/9901294},
	doi = {10.48550/ARXIV.ASTRO-PH/9901294},
	urldate = {2025-01-21},
	author = {Alibert, Y. and Baraffe, I. and Hauschildt, P. and Allard, F.},
	year = {1999},
	note = {Publisher: arXiv
Version Number: 1},
}

@article{ruoppo_improvement_2004,
	title = {Improvement of the {CORS} method for {Cepheids} radii determination based on {Strömgren} photometry},
	volume = {422},
	issn = {0004-6361, 1432-0746},
	url = {http://www.aanda.org/10.1051/0004-6361:20035901},
	doi = {10.1051/0004-6361:20035901},
	number = {1},
	urldate = {2025-01-21},
	journal = {Astronomy \& Astrophysics},
	author = {Ruoppo, A. and Ripepi, V. and Marconi, M. and Russo, G.},
	month = jul,
	year = {2004},
	pages = {253--265},
}

@article{rojo_arellano_new_1994,
	title = {A new approach to the surface brightness method for {Cepheid} radii determination.},
	volume = {29},
	issn = {0185-1101},
	url = {https://ui.adsabs.harvard.edu/abs/1994RMxAA..29..148R},
	urldate = {2025-01-22},
	journal = {Revista Mexicana de Astronomia y Astrofisica, vol. 29},
	author = {Rojo Arellano, E. and Arellano Ferro, A.},
	month = jul,
	year = {1994},
	note = {ADS Bibcode: 1994RMxAA..29..148R},
	pages = {148--152},
}

@article{petroni_classical_2003,
	title = {Classical {Cepheid} {Pulsation} {Models}. {IX}. {New} {Input} {Physics}},
	volume = {599},
	issn = {0004-637X, 1538-4357},
	url = {https://iopscience.iop.org/article/10.1086/379279},
	doi = {10.1086/379279},
	language = {en},
	number = {1},
	urldate = {2025-01-22},
	journal = {The Astrophysical Journal},
	author = {Petroni, Silvia and Bono, Giuseppe and Marconi, Marcella and Stellingwerf, Robert F.},
	month = dec,
	year = {2003},
	pages = {522--536},
}

@article{groenewegen_projection_2007,
	title = {The projection factor, period–radius relation, and surface–brightness colour relation in classical cepheids},
	volume = {474},
	issn = {0004-6361, 1432-0746},
	url = {http://www.aanda.org/10.1051/0004-6361:20078225},
	doi = {10.1051/0004-6361:20078225},
	number = {3},
	urldate = {2025-01-22},
	journal = {Astronomy \& Astrophysics},
	author = {Groenewegen, M. A. T.},
	month = nov,
	year = {2007},
	pages = {975--981},
}

@ARTICLE{groenewegen_pr_2013,
       author = {{Groenewegen}, M.~A.~T.},
        title = "{Baade-Wesselink distances to Galactic and Magellanic Cloud Cepheids and the effect of metallicity}",
      journal = {\aap},
         year = 2013,
        month = feb,
       volume = {550},
          eid = {A70},
        pages = {A70},
          doi = {10.1051/0004-6361/201220446},
archivePrefix = {arXiv},
       eprint = {1212.5478},
 primaryClass = {astro-ph.GA},
       adsurl = {https://ui.adsabs.harvard.edu/abs/2013A&A...550A..70G}
}

@article{turner_distance_2002,
	title = {The {Distance} {Scale} for {Classical} {Cepheid} {Variables}},
	volume = {124},
	issn = {00046256, 15383881},
	url = {https://iopscience.iop.org/article/10.1086/343774},
	doi = {10.1086/343774},
	number = {5},
	urldate = {2025-01-22},
	journal = {The Astronomical Journal},
	author = {Turner, David G. and Burke, James F.},
	month = nov,
	year = {2002},
	pages = {2931--2942},
}

@article{fernie_survey_1984,
	title = {A survey of {Cepheid} sizes},
	volume = {282},
	issn = {0004-637X, 1538-4357},
	url = {http://adsabs.harvard.edu/doi/10.1086/162243},
	doi = {10.1086/162243},
	language = {en},
	urldate = {2025-01-22},
	journal = {The Astrophysical Journal},
	author = {Fernie, J. D.},
	month = jul,
	year = {1984},
	pages = {641},
}

@article{cogan_radii_1978,
	title = {The radii and temperatures of classical {Cepheids}},
	volume = {221},
	issn = {0004-637X, 1538-4357},
	url = {http://adsabs.harvard.edu/doi/10.1086/156067},
	doi = {10.1086/156067},
	language = {en},
	urldate = {2025-01-22},
	journal = {The Astrophysical Journal},
	author = {Cogan, B. C.},
	month = apr,
	year = {1978},
	pages = {635},
}

@article{karp_hydrodynamic_1975,
	title = {Hydrodynamic models of a {Cepheid} atmosphere. {I} - {Deep} envelope models},
	volume = {199},
	issn = {0004-637X, 1538-4357},
	url = {http://adsabs.harvard.edu/doi/10.1086/153710},
	doi = {10.1086/153710},
	language = {en},
	urldate = {2025-01-22},
	journal = {The Astrophysical Journal},
	author = {Karp, A. H.},
	month = jul,
	year = {1975},
	pages = {448},
}

@article{laney_radii_1995,
	title = {The radii of {Galactic} {Cepheids}},
	volume = {274},
	issn = {0035-8711, 1365-2966},
	url = {https://academic.oup.com/mnras/article-lookup/doi/10.1093/mnras/274.2.337},
	doi = {10.1093/mnras/274.2.337},
	language = {en},
	number = {2},
	urldate = {2025-01-22},
	journal = {Monthly Notices of the Royal Astronomical Society},
	author = {Laney, C. D. and Stobie, R. S.},
	month = may,
	year = {1995},
	pages = {337--360},
}

@article{di_benedetto_pulsational_1994,
	title = {Pulsational parallaxes and calibration of the cosmic distance scale by {Cepheid} variable stars},
	volume = {285},
	issn = {0004-6361},
	url = {https://ui.adsabs.harvard.edu/abs/1994A&A...285..819D},
	urldate = {2025-01-22},
	journal = {Astronomy and Astrophysics},
	author = {di Benedetto, G. P.},
	month = may,
	year = {1994},
	note = {ADS Bibcode: 1994A\&A...285..819D},
	pages = {819--832},
}

@article{caccin_improvement_1981,
	title = {An improvement of the {Baade}-{Wesselink} method to determine the mean radius of pulsating variables.},
	volume = {97},
	issn = {0004-6361},
	url = {https://ui.adsabs.harvard.edu/abs/1981A&A....97..104C},
	urldate = {2025-01-22},
	journal = {Astronomy and Astrophysics},
	author = {Caccin, R. and Onnembo, A. and Russo, G. and Sollazzo, C.},
	month = apr,
	year = {1981},
	note = {ADS Bibcode: 1981A\&A....97..104C},
	pages = {104--109},
}

@misc{ripepi_cepheid_1996,
	title = {Cepheid {Radii} and the {Cors} {Method} {Revisited}},
	copyright = {Assumed arXiv.org perpetual, non-exclusive license to distribute this article for submissions made before January 2004},
	url = {https://arxiv.org/abs/astro-ph/9601019},
	doi = {10.48550/ARXIV.ASTRO-PH/9601019},
	urldate = {2025-01-22},
	publisher = {arXiv},
	author = {Ripepi, V. and Barone, F. and Milano, L. and Russo, G.},
	year = {1996},
	note = {Version Number: 1},
}

@ARTICLE{ripepi_1997_CORS,
       author = {{Ripepi}, V. and {Barone}, F. and {Milano}, L. and {Russo}, G.},
        title = "{Cepheid radii and the CORS method revisited.}",
      journal = {\aap},
         year = 1997,
        month = feb,
       volume = {318},
        pages = {797-804},
          doi = {10.48550/arXiv.astro-ph/9601019},
archivePrefix = {arXiv},
       eprint = {astro-ph/9601019},
 primaryClass = {astro-ph},
       adsurl = {https://ui.adsabs.harvard.edu/abs/1997A&A...318..797R}
}

@ARTICLE{ripepi_2019_pl,
       author = {{Ripepi}, V. and {Molinaro}, R. and {Musella}, I. and {Marconi}, M. and {Leccia}, S. and {Eyer}, L.},
        title = "{Reclassification of Cepheids in the Gaia Data Release 2. Period-luminosity and period-Wesenheit relations in the Gaia passbands}",
      journal = {\aap},
         year = 2019,
        month = may,
       volume = {625},
          eid = {A14},
        pages = {A14},
          doi = {10.1051/0004-6361/201834506},
archivePrefix = {arXiv},
       eprint = {1810.10486},
 primaryClass = {astro-ph.SR},
       adsurl = {https://ui.adsabs.harvard.edu/abs/2019A&A...625A..14R}
}

@ARTICLE{ripepi_2021,
       author = {{Ripepi}, V. and {Catanzaro}, G. and {Molinaro}, R. and {Gatto}, M. and {De Somma}, G. and {Marconi}, M. and {Romaniello}, M. and {Leccia}, S. and {Musella}, I. and {Trentin}, E. and {Clementini}, G. and {Testa}, V. and {Cusano}, F. and {Storm}, J.},
        title = "{Cepheid metallicity in the Leavitt law (C-metall) survey - I. HARPS-N@TNG spectroscopy of 47 classical Cepheids and 1 BL Her variables}",
      journal = {\mnras},
         year = 2021,
        month = dec,
       volume = {508},
       number = {3},
        pages = {4047-4071},
          doi = {10.1093/mnras/stab2460},
archivePrefix = {arXiv},
       eprint = {2108.11391},
 primaryClass = {astro-ph.GA},
       adsurl = {https://ui.adsabs.harvard.edu/abs/2021MNRAS.508.4047R}
}

@ARTICLE{ripepi_2022,
       author = {{Ripepi}, V. and {Catanzaro}, G. and {Clementini}, G. and {De Somma}, G. and {Drimmel}, R. and {Leccia}, S. and {Marconi}, M. and {Molinaro}, R. and {Musella}, I. and {Poggio}, E.},
        title = "{Classical Cepheid period-Wesenheit-metallicity relation in the Gaia bands}",
      journal = {\aap},
         year = 2022,
        month = mar,
       volume = {659},
          eid = {A167},
        pages = {A167},
          doi = {10.1051/0004-6361/202142649},
archivePrefix = {arXiv},
       eprint = {2201.01126},
 primaryClass = {astro-ph.SR},
       adsurl = {https://ui.adsabs.harvard.edu/abs/2022A&A...659A.167R}
}

@article{sollazzo_cepheid_1981,
	title = {Cepheid radii and masses by means of {VBLUW} photometry.},
	volume = {99},
	issn = {0004-6361},
	url = {https://ui.adsabs.harvard.edu/abs/1981A&A....99...66S},
	urldate = {2025-01-22},
	journal = {Astronomy and Astrophysics},
	author = {Sollazzo, C. and Russo, G. and Onnembo, A. and Caccin, B.},
	month = jun,
	year = {1981},
	note = {ADS Bibcode: 1981A\&A....99...66S},
	pages = {66--72},
}

@ARTICLE{gaia_collab_vallenari_2023_instru,
       author = {{Gaia Collaboration} and {Vallenari}, A. and {Brown}, A.~G.~A. and {Prusti}, T. and {de Bruijne}, J.~H.~J. and {Arenou}, F. and {Babusiaux}, C. and {Biermann}, M. and {Creevey}, O.~L. and {Ducourant}, C. and {Evans}, D.~W. and {Eyer}, L. and {Guerra}, R. and {Hutton}, A. and {Jordi}, C. and {Klioner}, S.~A. and {Lammers}, U.~L. and {Lindegren}, L. and {Luri}, X. and {Mignard}, F. and {Panem}, C. and {Pourbaix}, D. and {Randich}, S. and {Sartoretti}, P. and {Soubiran}, C. and {Tanga}, P. and {Walton}, N.~A. and {Bailer-Jones}, C.~A.~L. and {Bastian}, U. and {Drimmel}, R. and {Jansen}, F. and {Katz}, D. and {Lattanzi}, M.~G. and {van Leeuwen}, F. and {Bakker}, J. and {Cacciari}, C. and {Casta{\~n}eda}, J. and {De Angeli}, F. and {Fabricius}, C. and {Fouesneau}, M. and {Fr{\'e}mat}, Y. and {Galluccio}, L. and {Guerrier}, A. and {Heiter}, U. and {Masana}, E. and {Messineo}, R. and {Mowlavi}, N. and {Nicolas}, C. and {Nienartowicz}, K. and {Pailler}, F. and {Panuzzo}, P. and {Riclet}, F. and {Roux}, W. and {Seabroke}, G.~M. and {Sordo}, R. and {Th{\'e}venin}, F. and {Gracia-Abril}, G. and {Portell}, J. and {Teyssier}, D. and {Altmann}, M. and {Andrae}, R. and {Audard}, M. and {Bellas-Velidis}, I. and {Benson}, K. and {Berthier}, J. and {Blomme}, R. and {Burgess}, P.~W. and {Busonero}, D. and {Busso}, G. and {C{\'a}novas}, H. and {Carry}, B. and {Cellino}, A. and {Cheek}, N. and {Clementini}, G. and {Damerdji}, Y. and {Davidson}, M. and {de Teodoro}, P. and {Nu{\~n}ez Campos}, M. and {Delchambre}, L. and {Dell'Oro}, A. and {Esquej}, P. and {Fern{\'a}ndez-Hern{\'a}ndez}, J. and {Fraile}, E. and {Garabato}, D. and {Garc{\'\i}a-Lario}, P. and {Gosset}, E. and {Haigron}, R. and {Halbwachs}, J. -L. and {Hambly}, N.~C. and {Harrison}, D.~L. and {Hern{\'a}ndez}, J. and {Hestroffer}, D. and {Hodgkin}, S.~T. and {Holl}, B. and {Jan{\ss}en}, K. and {Jevardat de Fombelle}, G. and {Jordan}, S. and {Krone-Martins}, A. and {Lanzafame}, A.~C. and {L{\"o}ffler}, W. and {Marchal}, O. and {Marrese}, P.~M. and {Moitinho}, A. and {Muinonen}, K. and {Osborne}, P. and {Pancino}, E. and {Pauwels}, T. and {Recio-Blanco}, A. and {Reyl{\'e}}, C. and {Riello}, M. and {Rimoldini}, L. and {Roegiers}, T. and {Rybizki}, J. and {Sarro}, L.~M. and {Siopis}, C. and {Smith}, M. and {Sozzetti}, A. and {Utrilla}, E. and {van Leeuwen}, M. and {Abbas}, U. and {{\'A}brah{\'a}m}, P. and {Abreu Aramburu}, A. and {Aerts}, C. and {Aguado}, J.~J. and {Ajaj}, M. and {Aldea-Montero}, F. and {Altavilla}, G. and {{\'A}lvarez}, M.~A. and {Alves}, J. and {Anders}, F. and {Anderson}, R.~I. and {Anglada Varela}, E. and {Antoja}, T. and {Baines}, D. and {Baker}, S.~G. and {Balaguer-N{\'u}{\~n}ez}, L. and {Balbinot}, E. and {Balog}, Z. and {Barache}, C. and {Barbato}, D. and {Barros}, M. and {Barstow}, M.~A. and {Bartolom{\'e}}, S. and {Bassilana}, J. -L. and {Bauchet}, N. and {Becciani}, U. and {Bellazzini}, M. and {Berihuete}, A. and {Bernet}, M. and {Bertone}, S. and {Bianchi}, L. and {Binnenfeld}, A. and {Blanco-Cuaresma}, S. and {Blazere}, A. and {Boch}, T. and {Bombrun}, A. and {Bossini}, D. and {Bouquillon}, S. and {Bragaglia}, A. and {Bramante}, L. and {Breedt}, E. and {Bressan}, A. and {Brouillet}, N. and {Brugaletta}, E. and {Bucciarelli}, B. and {Burlacu}, A. and {Butkevich}, A.~G. and {Buzzi}, R. and {Caffau}, E. and {Cancelliere}, R. and {Cantat-Gaudin}, T. and {Carballo}, R. and {Carlucci}, T. and {Carnerero}, M.~I. and {Carrasco}, J.~M. and {Casamiquela}, L. and {Castellani}, M. and {Castro-Ginard}, A. and {Chaoul}, L. and {Charlot}, P. and {Chemin}, L. and {Chiaramida}, V. and {Chiavassa}, A. and {Chornay}, N. and {Comoretto}, G. and {Contursi}, G. and {Cooper}, W.~J. and {Cornez}, T. and {Cowell}, S. and {Crifo}, F. and {Cropper}, M. and {Crosta}, M. and {Crowley}, C. and {Dafonte}, C. and {Dapergolas}, A. and {David}, M. and {David}, P. and {de Laverny}, P. and {De Luise}, F. and {De March}, R.},
        title = "{Gaia Data Release 3. Summary of the content and survey properties}",
      journal = {\aap},
         year = 2023,
        month = jun,
       volume = {674},
          eid = {A1},
        pages = {A1},
          doi = {10.1051/0004-6361/202243940},
archivePrefix = {arXiv},
       eprint = {2208.00211},
 primaryClass = {astro-ph.GA},
       adsurl = {https://ui.adsabs.harvard.edu/abs/2023A&A...674A...1G}
}

@dataset{gaia_collab_2022_variability,
       author = {{Gaia Collaboration}},
        title = "{VizieR Online Data Catalog: Gaia DR3 Part 4. Variability (Gaia Collaboration, 2022)}",
 howpublished = {VizieR On-line Data Catalog: I/358.  Originally published in: 2023A\&A...674A...1G},
         year = 2022,
        month = may,
          eid = {I/358},
       adsurl = {https://ui.adsabs.harvard.edu/abs/2022yCat.1358....0G}
}

@article{gaia_collaboration_gaia_2023_phot,
	title = {\textit{{Gaia}} {Data} {Release} 3: {The} {Galaxy} in your preferred colours: {Synthetic} photometry from \textit{{Gaia}} low-resolution spectra},
	volume = {674},
	copyright = {https://creativecommons.org/licenses/by/4.0},
	issn = {0004-6361, 1432-0746},
	shorttitle = {\textit{{Gaia}} {Data} {Release} 3},
	url = {https://www.aanda.org/10.1051/0004-6361/202243709},
	doi = {10.1051/0004-6361/202243709},
	language = {en},
	urldate = {2025-02-04},
	journal = {Astronomy \& Astrophysics},
	author = {{Gaia Collaboration} and Montegriffo, P. and Bellazzini, M. and De Angeli, F. and Andrae, R. and Barstow, M. A. and Bossini, D. and Bragaglia, A. and Burgess, P. W. and Cacciari, C. and Carrasco, J. M. and Chornay, N. and Delchambre, L. and Evans, D. W. and Fouesneau, M. and Frémat, Y. and Garabato, D. and Jordi, C. and Manteiga, M. and Massari, D. and Palaversa, L. and Pancino, E. and Riello, M. and Ruz Mieres, D. and Sanna, N. and Santoveña, R. and Sordo, R. and Vallenari, A. and Walton, N. A. and Brown, A. G. A. and Prusti, T. and De Bruijne, J. H. J. and Arenou, F. and Babusiaux, C. and Biermann, M. and Creevey, O. L. and Ducourant, C. and Eyer, L. and Guerra, R. and Hutton, A. and Klioner, S. A. and Lammers, U. L. and Lindegren, L. and Luri, X. and Mignard, F. and Panem, C. and Pourbaix, D. and Randich, S. and Sartoretti, P. and Soubiran, C. and Tanga, P. and Bailer-Jones, C. A. L. and Bastian, U. and Drimmel, R. and Jansen, F. and Katz, D. and Lattanzi, M. G. and Van Leeuwen, F. and Bakker, J. and Castañeda, J. and Fabricius, C. and Galluccio, L. and Guerrier, A. and Heiter, U. and Masana, E. and Messineo, R. and Mowlavi, N. and Nicolas, C. and Nienartowicz, K. and Pailler, F. and Panuzzo, P. and Riclet, F. and Roux, W. and Seabroke, G. M. and Thévenin, F. and Gracia-Abril, G. and Portell, J. and Teyssier, D. and Altmann, M. and Audard, M. and Bellas-Velidis, I. and Benson, K. and Berthier, J. and Blomme, R. and Busonero, D. and Busso, G. and Cánovas, H. and Carry, B. and Cellino, A. and Cheek, N. and Clementini, G. and Damerdji, Y. and Davidson, M. and De Teodoro, P. and Nuñez Campos, M. and Dell’Oro, A. and Esquej, P. and Fernández-Hernández, J. and Fraile, E. and García-Lario, P. and Gosset, E. and Haigron, R. and Halbwachs, J.-L. and Hambly, N. C. and Harrison, D. L. and Hernández, J. and Hestroffer, D. and Hodgkin, S. T. and Holl, B. and Janßen, K. and Jevardat De Fombelle, G. and Jordan, S. and Krone-Martins, A. and Lanzafame, A. C. and Löffler, W. and Marchal, O. and Marrese, P. M. and Moitinho, A. and Muinonen, K. and Osborne, P. and Pauwels, T. and Recio-Blanco, A. and Reylé, C. and Rimoldini, L. and Roegiers, T. and Rybizki, J. and Sarro, L. M. and Siopis, C. and Smith, M. and Sozzetti, A. and Utrilla, E. and Van Leeuwen, M. and Abbas, U. and Ábrahám, P. and Abreu Aramburu, A. and Aerts, C. and Aguado, J. J. and Ajaj, M. and Aldea-Montero, F. and Altavilla, G. and Álvarez, M. A. and Alves, J. and Anderson, R. I. and Anglada Varela, E. and Antoja, T. and Baines, D. and Baker, S. G. and Balaguer-Núñez, L. and Balbinot, E. and Balog, Z. and Barache, C. and Barbato, D. and Barros, M. and Bartolomé, S. and Bassilana, J.-L. and Bauchet, N. and Becciani, U. and Berihuete, A. and Bernet, M. and Bertone, S. and Bianchi, L. and Binnenfeld, A. and Blanco-Cuaresma, S. and Boch, T. and Bombrun, A. and Bouquillon, S. and Bramante, L. and Breedt, E. and Bressan, A. and Brouillet, N. and Brugaletta, E. and Bucciarelli, B. and Burlacu, A. and Butkevich, A. G. and Buzzi, R. and Caffau, E. and Cancelliere, R. and Cantat-Gaudin, T. and Carballo, R. and Carlucci, T. and Carnerero, M. I. and Casamiquela, L. and Castellani, M. and Castro-Ginard, A. and Chaoul, L. and Charlot, P. and Chemin, L. and Chiaramida, V. and Chiavassa, A. and Comoretto, G. and Contursi, G. and Cooper, W. J. and Cornez, T. and Cowell, S. and Crifo, F. and Cropper, M. and Crosta, M. and Crowley, C. and Dafonte, C. and Dapergolas, A. and David, P. and De Laverny, P. and De Luise, F. and De March, R. and De Ridder, J. and De Souza, R. and De Torres, A. and Del Peloso, E. F. and Del Pozo, E. and Delbo, M. and Delgado, A. and Delisle, J.-B. and Demouchy, C. and Dharmawardena, T. E. and Diakite, S. and Diener, C. and Distefano, E. and Dolding, C. and Enke, H. and Fabre, C. and Fabrizio, M. and Faigler, S. and Fedorets, G. and Fernique, P. and Figueras, F. and Fournier, Y. and Fouron, C. and Fragkoudi, F. and Gai, M. and Garcia-Gutierrez, A. and Garcia-Reinaldos, M. and García-Torres, M. and Garofalo, A. and Gavel, A. and Gavras, P. and Gerlach, E. and Geyer, R. and Giacobbe, P. and Gilmore, G. and Girona, S. and Giuffrida, G. and Gomel, R. and Gomez, A. and González-Núñez, J. and González-Santamaría, I. and González-Vidal, J. J. and Granvik, M. and Guillout, P. and Guiraud, J. and Gutiérrez-Sánchez, R. and Guy, L. P. and Hatzidimitriou, D. and Hauser, M. and Haywood, M. and Helmer, A. and Helmi, A. and Sarmiento, M. H. and Hidalgo, S. L. and Hładczuk, N. and Hobbs, D. and Holland, G. and Huckle, H. E. and Jardine, K. and Jasniewicz, G. and Jean-Antoine Piccolo, A. and Jiménez-Arranz, Ó. and Juaristi Campillo, J. and Julbe, F. and Karbevska, L. and Kervella, P. and Khanna, S. and Kordopatis, G. and Korn, A. J. and Kóspál, Á and Kostrzewa-Rutkowska, Z. and Kruszyńska, K. and Kun, M. and Laizeau, P. and Lambert, S. and Lanza, A. F. and Lasne, Y. and Le Campion, J.-F. and Lebreton, Y. and Lebzelter, T. and Leccia, S. and Leclerc, N. and Lecoeur-Taibi, I. and Liao, S. and Licata, E. L. and Lindstróm, H. E. P. and Lister, T. A. and Livanou, E. and Lobel, A. and Lorca, A. and Loup, C. and Madrero Pardo, P. and Magdaleno Romeo, A. and Managau, S. and Mann, R. G. and Marchant, J. M. and Marconi, M. and Marcos, J. and Marcos Santos, M. M. S. and Marín Pina, D. and Marinoni, S. and Marocco, F. and Marshall, D. J. and Martin Polo, L. and Martín-Fleitas, J. M. and Marton, G. and Mary, N. and Masip, A. and Mastrobuono-Battisti, A. and Mazeh, T. and McMillan, P. J. and Messina, S. and Michalik, D. and Millar, N. R. and Mints, A. and Molina, D. and Molinaro, R. and Molnár, L. and Monari, G. and Monguió, M. and Montero, A. and Mor, R. and Mora, A. and Morbidelli, R. and Morel, T. and Morris, D. and Muraveva, T. and Murphy, C. P. and Musella, I. and Nagy, Z. and Noval, L. and Ocaña, F. and Ogden, A. and Ordenovic, C. and Osinde, J. O. and Pagani, C. and Pagano, I. and Palicio, P. A. and Pallas-Quintela, L. and Panahi, A. and Payne-Wardenaar, S. and Peñalosa Esteller, X. and Penttilä, A. and Pichon, B. and Piersimoni, A. M. and Pineau, F.-X. and Plachy, E. and Plum, G. and Poggio, E. and Prša, A. and Pulone, L. and Racero, E. and Ragaini, S. and Rainer, M. and Raiteri, C. M. and Ramos, P. and Ramos-Lerate, M. and Re Fiorentin, P. and Regibo, S. and Richards, P. J. and Rios Diaz, C. and Ripepi, V. and Riva, A. and Rix, H.-W. and Rixon, G. and Robichon, N. and Robin, A. C. and Robin, C. and Roelens, M. and Rogues, H. R. O. and Rohrbasser, L. and Romero-Gómez, M. and Rowell, N. and Royer, F. and Rybicki, K. A. and Sadowski, G. and Sáez Núñez, A. and Sagristà Sellés, A. and Sahlmann, J. and Salguero, E. and Samaras, N. and Sanchez Gimenez, V. and Sarasso, M. and Schultheis, M. S. and Sciacca, E. and Segol, M. and Segovia, J. C. and Ségransan, D. and Semeux, D. and Shahaf, S. and Siddiqui, H. I. and Siebert, A. and Siltala, L. and Silvelo, A. and Slezak, E. and Slezak, I. and Smart, R. L. and Snaith, O. N. and Solano, E. and Solitro, F. and Souami, D. and Souchay, J. and Spagna, A. and Spina, L. and Spoto, F. and Steele, I. A. and Steidelmüller, H. and Stephenson, C. A. and Süveges, M. and Surdej, J. and Szabados, L. and Szegedi-Elek, E. and Taris, F. and Taylor, M. B. and Teixeira, R. and Tolomei, L. and Tonello, N. and Torra, F. and Torra, J. and Torralba Elipe, G. and Trabucchi, M. and Tsounis, A. T. and Turon, C. and Ulla, A. and Unger, N. and Vaillant, M. V. and Van Dillen, E. and Van Reeven, W. and Vanel, O. and Vecchiato, A. and Viala, Y. and Vicente, D. and Voutsinas, S. and Wevers, T. and Wyrzykowski, Ł. and Yoldas, A. and Yvard, P. and Zhao, H. and Zorec, J. and Zucker, S. and Zwitter, T.},
	month = jun,
	year = {2023},
	pages = {A33},
}

@ARTICLE{recioBlanco_2023,
       author = {{Recio-Blanco}, A. and {de Laverny}, P. and {Palicio}, P.~A. and {Kordopatis}, G. and {{\'A}lvarez}, M.~A. and {Schultheis}, M. and {Contursi}, G. and {Zhao}, H. and {Torralba Elipe}, G. and {Ordenovic}, C. and {Manteiga}, M. and {Dafonte}, C. and {Oreshina-Slezak}, I. and {Bijaoui}, A. and {Fr{\'e}mat}, Y. and {Seabroke}, G. and {Pailler}, F. and {Spitoni}, E. and {Poggio}, E. and {Creevey}, O.~L. and {Abreu Aramburu}, A. and {Accart}, S. and {Andrae}, R. and {Bailer-Jones}, C.~A.~L. and {Bellas-Velidis}, I. and {Brouillet}, N. and {Brugaletta}, E. and {Burlacu}, A. and {Carballo}, R. and {Casamiquela}, L. and {Chiavassa}, A. and {Cooper}, W.~J. and {Dapergolas}, A. and {Delchambre}, L. and {Dharmawardena}, T.~E. and {Drimmel}, R. and {Edvardsson}, B. and {Fouesneau}, M. and {Garabato}, D. and {Garc{\'\i}a-Lario}, P. and {Garc{\'\i}a-Torres}, M. and {Gavel}, A. and {Gomez}, A. and {Gonz{\'a}lez-Santamar{\'\i}a}, I. and {Hatzidimitriou}, D. and {Heiter}, U. and {Jean-Antoine Piccolo}, A. and {Kontizas}, M. and {Korn}, A.~J. and {Lanzafame}, A.~C. and {Lebreton}, Y. and {Le Fustec}, Y. and {Licata}, E.~L. and {Lindstr{\o}m}, H.~E.~P. and {Livanou}, E. and {Lobel}, A. and {Lorca}, A. and {Magdaleno Romeo}, A. and {Marocco}, F. and {Marshall}, D.~J. and {Mary}, N. and {Nicolas}, C. and {Pallas-Quintela}, L. and {Panem}, C. and {Pichon}, B. and {Riclet}, F. and {Robin}, C. and {Rybizki}, J. and {Santove{\~n}a}, R. and {Silvelo}, A. and {Smart}, R.~L. and {Sarro}, L.~M. and {Sordo}, R. and {Soubiran}, C. and {S{\"u}veges}, M. and {Ulla}, A. and {Vallenari}, A. and {Zorec}, J. and {Utrilla}, E. and {Bakker}, J.},
        title = "{Gaia Data Release 3. Analysis of RVS spectra using the General Stellar Parametriser from spectroscopy}",
      journal = {\aap},
         year = 2023,
        month = jun,
       volume = {674},
          eid = {A29},
        pages = {A29},
          doi = {10.1051/0004-6361/202243750},
archivePrefix = {arXiv},
       eprint = {2206.05541},
 primaryClass = {astro-ph.GA},
       adsurl = {https://ui.adsabs.harvard.edu/abs/2023A&A...674A..29R}
}

@ARTICLE{bailer_2021,
       author = {{Bailer-Jones}, C.~A.~L. and {Rybizki}, J. and {Fouesneau}, M. and {Demleitner}, M. and {Andrae}, R.},
        title = "{Estimating Distances from Parallaxes. V. Geometric and Photogeometric Distances to 1.47 Billion Stars in Gaia Early Data Release 3}",
      journal = {\aj},
         year = 2021,
        month = mar,
       volume = {161},
       number = {3},
          eid = {147},
        pages = {147},
          doi = {10.3847/1538-3881/abd806},
archivePrefix = {arXiv},
       eprint = {2012.05220},
 primaryClass = {astro-ph.SR},
       adsurl = {https://ui.adsabs.harvard.edu/abs/2021AJ....161..147B}
}

@ARTICLE{lindegren_2021one,
       author = {{Lindegren}, L. and {Klioner}, S.~A. and {Hern{\'a}ndez}, J. and {Bombrun}, A. and {Ramos-Lerate}, M. and {Steidelm{\"u}ller}, H. and {Bastian}, U. and {Biermann}, M. and {de Torres}, A. and {Gerlach}, E. and {Geyer}, R. and {Hilger}, T. and {Hobbs}, D. and {Lammers}, U. and {McMillan}, P.~J. and {Stephenson}, C.~A. and {Casta{\~n}eda}, J. and {Davidson}, M. and {Fabricius}, C. and {Gracia-Abril}, G. and {Portell}, J. and {Rowell}, N. and {Teyssier}, D. and {Torra}, F. and {Bartolom{\'e}}, S. and {Clotet}, M. and {Garralda}, N. and {Gonz{\'a}lez-Vidal}, J.~J. and {Torra}, J. and {Abbas}, U. and {Altmann}, M. and {Anglada Varela}, E. and {Balaguer-N{\'u}{\~n}ez}, L. and {Balog}, Z. and {Barache}, C. and {Becciani}, U. and {Bernet}, M. and {Bertone}, S. and {Bianchi}, L. and {Bouquillon}, S. and {Brown}, A.~G.~A. and {Bucciarelli}, B. and {Busonero}, D. and {Butkevich}, A.~G. and {Buzzi}, R. and {Cancelliere}, R. and {Carlucci}, T. and {Charlot}, P. and {Cioni}, M. -R.~L. and {Crosta}, M. and {Crowley}, C. and {del Peloso}, E.~F. and {del Pozo}, E. and {Drimmel}, R. and {Esquej}, P. and {Fienga}, A. and {Fraile}, E. and {Gai}, M. and {Garcia-Reinaldos}, M. and {Guerra}, R. and {Hambly}, N.~C. and {Hauser}, M. and {Jan{\ss}en}, K. and {Jordan}, S. and {Kostrzewa-Rutkowska}, Z. and {Lattanzi}, M.~G. and {Liao}, S. and {Licata}, E. and {Lister}, T.~A. and {L{\"o}ffler}, W. and {Marchant}, J.~M. and {Masip}, A. and {Mignard}, F. and {Mints}, A. and {Molina}, D. and {Mora}, A. and {Morbidelli}, R. and {Murphy}, C.~P. and {Pagani}, C. and {Panuzzo}, P. and {Pe{\~n}alosa Esteller}, X. and {Poggio}, E. and {Re Fiorentin}, P. and {Riva}, A. and {Sagrist{\`a} Sell{\'e}s}, A. and {Sanchez Gimenez}, V. and {Sarasso}, M. and {Sciacca}, E. and {Siddiqui}, H.~I. and {Smart}, R.~L. and {Souami}, D. and {Spagna}, A. and {Steele}, I.~A. and {Taris}, F. and {Utrilla}, E. and {van Reeven}, W. and {Vecchiato}, A.},
        title = "{Gaia Early Data Release 3. The astrometric solution}",
      journal = {\aap},
         year = 2021,
        month = may,
       volume = {649},
          eid = {A2},
        pages = {A2},
          doi = {10.1051/0004-6361/202039709},
archivePrefix = {arXiv},
       eprint = {2012.03380},
 primaryClass = {astro-ph.IM},
       adsurl = {https://ui.adsabs.harvard.edu/abs/2021A&A...649A...2L}
}

@ARTICLE{wesselink_1969,
       author = {{Wesselink}, A.~J.},
        title = "{Surface brightnesses in the U, B, V system with applications of M\_{\ensuremath{\upsilon}} and dimensions of stars}",
      journal = {\mnras},
         year = 1969,
        month = jan,
       volume = {144},
        pages = {297},
          doi = {10.1093/mnras/144.3.297},
       adsurl = {https://ui.adsabs.harvard.edu/abs/1969MNRAS.144..297W}
}

@ARTICLE{barnes_1976,
       author = {{Barnes}, T.~G. and {Evans}, D.~S.},
        title = "{Stellar angular diameters and visual surface brightness - I. Late spectral types.}",
      journal = {\mnras},
         year = 1976,
        month = mar,
       volume = {174},
        pages = {489-502},
          doi = {10.1093/mnras/174.3.489},
       adsurl = {https://ui.adsabs.harvard.edu/abs/1976MNRAS.174..489B}
}

@ARTICLE{mamajek_2015,
       author = {{Mamajek}, E.~E. and {Torres}, G. and {Prsa}, A. and {Harmanec}, P. and {Asplund}, M. and {Bennett}, P.~D. and {Capitaine}, N. and {Christensen-Dalsgaard}, J. and {Depagne}, E. and {Folkner}, W.~M. and {Haberreiter}, M. and {Hekker}, S. and {Hilton}, J.~L. and {Kostov}, V. and {Kurtz}, D.~W. and {Laskar}, J. and {Mason}, B.~D. and {Milone}, E.~F. and {Montgomery}, M.~M. and {Richards}, M.~T. and {Schou}, J. and {Stewart}, S.~G.},
        title = "{IAU 2015 Resolution B2 on Recommended Zero Points for the Absolute and Apparent Bolometric Magnitude Scales}",
      journal = {arXiv e-prints},
         year = 2015,
        month = oct,
          eid = {arXiv:1510.06262},
        pages = {arXiv:1510.06262},
          doi = {10.48550/arXiv.1510.06262},
archivePrefix = {arXiv},
       eprint = {1510.06262},
 primaryClass = {astro-ph.SR},
       adsurl = {https://ui.adsabs.harvard.edu/abs/2015arXiv151006262M}
}

@ARTICLE{fouque_1997,
       author = {{Fouque}, P. and {Gieren}, W.~P.},
        title = "{An improved calibration of Cepheid visual and infrared surface brightness relations from accurate angular diameter measurements of cool giants and supergiants.}",
      journal = {\aap},
         year = 1997,
        month = apr,
       volume = {320},
        pages = {799-810},
       adsurl = {https://ui.adsabs.harvard.edu/abs/1997A&A...320..799F}
}

@ARTICLE{pietrzynski_2013,
       author = {{Pietrzy{\'n}ski}, G. and {Graczyk}, D. and {Gieren}, W. and {Thompson}, I.~B. and {Pilecki}, B. and {Udalski}, A. and {Soszy{\'n}ski}, I. and {Koz{\l}owski}, S. and {Konorski}, P. and {Suchomska}, K. and {Bono}, G. and {Moroni}, P.~G. Prada and {Villanova}, S. and {Nardetto}, N. and {Bresolin}, F. and {Kudritzki}, R.~P. and {Storm}, J. and {Gallenne}, A. and {Smolec}, R. and {Minniti}, D. and {Kubiak}, M. and {Szyma{\'n}ski}, M.~K. and {Poleski}, R. and {Wyrzykowski}, {\L}. and {Ulaczyk}, K. and {Pietrukowicz}, P. and {G{\'o}rski}, M. and {Karczmarek}, P.},
        title = "{An eclipsing-binary distance to the Large Magellanic Cloud accurate to two per cent}",
      journal = {\nat},
         year = 2013,
        month = mar,
       volume = {495},
       number = {7439},
        pages = {76-79},
          doi = {10.1038/nature11878},
archivePrefix = {arXiv},
       eprint = {1303.2063},
 primaryClass = {astro-ph.GA},
       adsurl = {https://ui.adsabs.harvard.edu/abs/2013Natur.495...76P}
}

@ARTICLE{pietrzynski_2019,
       author = {{Pietrzy{\'n}ski}, G. and {Graczyk}, D. and {Gallenne}, A. and {Gieren}, W. and {Thompson}, I.~B. and {Pilecki}, B. and {Karczmarek}, P. and {G{\'o}rski}, M. and {Suchomska}, K. and {Taormina}, M. and {Zgirski}, B. and {Wielg{\'o}rski}, P. and {Ko{\l}aczkowski}, Z. and {Konorski}, P. and {Villanova}, S. and {Nardetto}, N. and {Kervella}, P. and {Bresolin}, F. and {Kudritzki}, R.~P. and {Storm}, J. and {Smolec}, R. and {Narloch}, W.},
        title = "{A distance to the Large Magellanic Cloud that is precise to one per cent}",
      journal = {\nat},
         year = 2019,
        month = mar,
       volume = {567},
       number = {7747},
        pages = {200-203},
          doi = {10.1038/s41586-019-0999-4},
archivePrefix = {arXiv},
       eprint = {1903.08096},
 primaryClass = {astro-ph.GA},
       adsurl = {https://ui.adsabs.harvard.edu/abs/2019Natur.567..200P}
}

@ARTICLE{graczyk_2020,
       author = {{Graczyk}, Dariusz and {Pietrzy{\'n}ski}, Grzegorz and {Thompson}, Ian B. and {Gieren}, Wolfgang and {Zgirski}, Bart{\l}omiej and {Villanova}, Sandro and {G{\'o}rski}, Marek and {Wielg{\'o}rski}, Piotr and {Karczmarek}, Paulina and {Narloch}, Weronika and {Pilecki}, Bogumi{\l} and {Taormina}, Monica and {Smolec}, Rados{\l}aw and {Suchomska}, Ksenia and {Gallenne}, Alexandre and {Nardetto}, Nicolas and {Storm}, Jesper and {Kudritzki}, Rolf-Peter and {Ka{\l}uszy{\'n}ski}, Miko{\l}aj and {Pych}, Wojciech},
        title = "{A Distance Determination to the Small Magellanic Cloud with an Accuracy of Better than Two Percent Based on Late-type Eclipsing Binary Stars}",
      journal = {\apj},
         year = 2020,
        month = nov,
       volume = {904},
       number = {1},
          eid = {13},
        pages = {13},
          doi = {10.3847/1538-4357/abbb2b},
archivePrefix = {arXiv},
       eprint = {2010.08754},
 primaryClass = {astro-ph.GA},
       adsurl = {https://ui.adsabs.harvard.edu/abs/2020ApJ...904...13G}
}

@ARTICLE{riello_2021,
       author = {{Riello}, M. and {De Angeli}, F. and {Evans}, D.~W. and {Montegriffo}, P. and {Carrasco}, J.~M. and {Busso}, G. and {Palaversa}, L. and {Burgess}, P.~W. and {Diener}, C. and {Davidson}, M. and {Rowell}, N. and {Fabricius}, C. and {Jordi}, C. and {Bellazzini}, M. and {Pancino}, E. and {Harrison}, D.~L. and {Cacciari}, C. and {van Leeuwen}, F. and {Hambly}, N.~C. and {Hodgkin}, S.~T. and {Osborne}, P.~J. and {Altavilla}, G. and {Barstow}, M.~A. and {Brown}, A.~G.~A. and {Castellani}, M. and {Cowell}, S. and {De Luise}, F. and {Gilmore}, G. and {Giuffrida}, G. and {Hidalgo}, S. and {Holland}, G. and {Marinoni}, S. and {Pagani}, C. and {Piersimoni}, A.~M. and {Pulone}, L. and {Ragaini}, S. and {Rainer}, M. and {Richards}, P.~J. and {Sanna}, N. and {Walton}, N.~A. and {Weiler}, M. and {Yoldas}, A.},
        title = "{Gaia Early Data Release 3. Photometric content and validation}",
      journal = {\aap},
         year = 2021,
        month = may,
       volume = {649},
          eid = {A3},
        pages = {A3},
          doi = {10.1051/0004-6361/202039587},
archivePrefix = {arXiv},
       eprint = {2012.01916},
 primaryClass = {astro-ph.IM},
       adsurl = {https://ui.adsabs.harvard.edu/abs/2021A&A...649A...3R}
}

@ARTICLE{lallement_2022,
       author = {{Lallement}, R. and {Vergely}, J.~L. and {Babusiaux}, C. and {Cox}, N.~L.~J.},
        title = "{Updated Gaia-2MASS 3D maps of Galactic interstellar dust}",
      journal = {\aap},
         year = 2022,
        month = may,
       volume = {661},
          eid = {A147},
        pages = {A147},
          doi = {10.1051/0004-6361/202142846},
archivePrefix = {arXiv},
       eprint = {2203.01627},
 primaryClass = {astro-ph.GA},
       adsurl = {https://ui.adsabs.harvard.edu/abs/2022A&A...661A.147L}
}

@ARTICLE{hocde_2023,
       author = {{Hocd{\'e}}, V. and {Smolec}, R. and {Moskalik}, P. and {Zi{\'o}{\l}kowska}, O. and {Singh Rathour}, R.},
        title = "{Metallicity estimations of MW, SMC, and LMC classical Cepheids from the shape of the V- and I-band light curves}",
      journal = {\aap},
         year = 2023,
        month = mar,
       volume = {671},
          eid = {A157},
        pages = {A157},
          doi = {10.1051/0004-6361/202245038},
archivePrefix = {arXiv},
       eprint = {2301.00229},
 primaryClass = {astro-ph.SR},
       adsurl = {https://ui.adsabs.harvard.edu/abs/2023A&A...671A.157H}
}

@ARTICLE{luck_2018,
       author = {{Luck}, R. Earle},
        title = "{Cepheid Abundances: Multiphase Results and Spatial Gradients}",
      journal = {\aj},
         year = 2018,
        month = oct,
       volume = {156},
       number = {4},
          eid = {171},
        pages = {171},
          doi = {10.3847/1538-3881/aadcac},
archivePrefix = {arXiv},
       eprint = {1808.05863},
 primaryClass = {astro-ph.SR},
       adsurl = {https://ui.adsabs.harvard.edu/abs/2018AJ....156..171L}
}

@misc{bailleul_2025,
      title={Surface brightness-colour relations of Cepheids calibrated by optical interferometry}, 
      author={M. C. Bailleul and N. Nardetto and V. Hocdé and P. Kervella and W. Gieren and J. Storm and G. Pietrzyński and A. Gallenne and A. Mérand and G. Bras and A. Recio Blanco and P. de Laverny and P. A. Palacio and A. Afanasiev and W. Kiviaho},
      year={2025},
      eprint={2504.13581},
      archivePrefix={arXiv},
      primaryClass={astro-ph.SR},
      url={https://arxiv.org/abs/2504.13581}, 
}

@ARTICLE{nardetto_2007,
       author = {{Nardetto}, N. and {Mourard}, D. and {Mathias}, Ph. and {Fokin}, A. and {Gillet}, D.},
        title = "{High-resolution spectroscopy for Cepheids distance determination. II. A period-projection factor relation}",
      journal = {\aap},
         year = 2007,
        month = aug,
       volume = {471},
       number = {2},
        pages = {661-669},
          doi = {10.1051/0004-6361:20066853},
archivePrefix = {arXiv},
       eprint = {0804.1330},
 primaryClass = {astro-ph},
       adsurl = {https://ui.adsabs.harvard.edu/abs/2007A&A...471..661N}
}

@ARTICLE{kervella_2004_pr_relation,
       author = {{Kervella}, P. and {Bersier}, D. and {Mourard}, D. and {Nardetto}, N. and {Coud{\'e} du Foresto}, V.},
        title = "{Cepheid distances from infrared long-baseline interferometry. II. Calibration of the period-radius and period-luminosity relations}",
      journal = {\aap},
         year = 2004,
        month = aug,
       volume = {423},
        pages = {327-333},
          doi = {10.1051/0004-6361:20035596},
archivePrefix = {arXiv},
       eprint = {astro-ph/0404179},
 primaryClass = {astro-ph},
       adsurl = {https://ui.adsabs.harvard.edu/abs/2004A&A...423..327K}
}

@dataset{gaia_EDR3_2020,
       author = {{Gaia Collaboration}},
        title = "{VizieR Online Data Catalog: Gaia EDR3 (Gaia Collaboration, 2020)}",
 howpublished = {VizieR On-line Data Catalog: I/350.  Originally published in: 2021A\&A...649A...1G},
         year = 2020,
        month = nov,
          eid = {I/350},
          doi = {10.26093/cds/vizier.1350},
       adsurl = {https://ui.adsabs.harvard.edu/abs/2020yCat.1350....0G}
}

@ARTICLE{Skowron_2019,
       author = {{Skowron}, Dorota M. and {Skowron}, Jan and {Mr{\'o}z}, Przemek and {Udalski}, Andrzej and {Pietrukowicz}, Pawe{\l} and {Soszy{\'n}ski}, Igor and {Szyma{\'n}ski}, Micha{\l} K. and {Poleski}, Rados{\l}aw and {Koz{\l}owski}, Szymon and {Ulaczyk}, Krzysztof and {Rybicki}, Krzysztof and {Iwanek}, Patryk},
        title = "{A three-dimensional map of the Milky Way using classical Cepheid variable stars}",
      journal = {Science},
         year = 2019,
        month = aug,
       volume = {365},
       number = {6452},
        pages = {478-482},
          doi = {10.1126/science.aau3181},
archivePrefix = {arXiv},
       eprint = {1806.10653},
 primaryClass = {astro-ph.GA},
       adsurl = {https://ui.adsabs.harvard.edu/abs/2019Sci...365..478S}
}

@ARTICLE{kervella_2024_rsPup,
       author = {{Kervella}, P. and {Bond}, H.~E. and {Cracraft}, M. and {Szabados}, L. and {Breitfelder}, J. and {M{\'e}rand}, A. and {Sparks}, W.~B. and {Gallenne}, A. and {Bersier}, D. and {Fouqu{\'e}}, P. and {Anderson}, R.~I.},
        title = "{The long-period Galactic Cepheid RS Puppis. III. A geometric distance from HST polarimetric imaging of its light echoes}",
      journal = {\aap},
         year = 2014,
        month = dec,
       volume = {572},
          eid = {A7},
        pages = {A7},
          doi = {10.1051/0004-6361/201424395},
archivePrefix = {arXiv},
       eprint = {1408.1697},
 primaryClass = {astro-ph.SR},
       adsurl = {https://ui.adsabs.harvard.edu/abs/2014A&A...572A...7K}
}

@ARTICLE{luck_2011,
       author = {{Luck}, R.~E. and {Andrievsky}, S.~M. and {Kovtyukh}, V.~V. and {Gieren}, W. and {Graczyk}, D.},
        title = "{The Distribution of the Elements in the Galactic Disk. II. Azimuthal and Radial Variation in Abundances from Cepheids}",
      journal = {\aj},
         year = 2011,
        month = aug,
       volume = {142},
       number = {2},
          eid = {51},
        pages = {51},
          doi = {10.1088/0004-6256/142/2/51},
archivePrefix = {arXiv},
       eprint = {1106.0182},
 primaryClass = {astro-ph.GA},
       adsurl = {https://ui.adsabs.harvard.edu/abs/2011AJ....142...51L}
}

@ARTICLE{kervella_2004_I_distances,
       author = {{Kervella}, P. and {Nardetto}, N. and {Bersier}, D. and {Mourard}, D. and {Coud{\'e} du Foresto}, V.},
        title = "{Cepheid distances from infrared long-baseline interferometry. I. VINCI/VLTI observations of seven Galactic Cepheids}",
      journal = {\aap},
         year = 2004,
        month = mar,
       volume = {416},
        pages = {941-953},
          doi = {10.1051/0004-6361:20031743},
archivePrefix = {arXiv},
       eprint = {astro-ph/0311525},
 primaryClass = {astro-ph},
       adsurl = {https://ui.adsabs.harvard.edu/abs/2004A&A...416..941K}
}

@ARTICLE{kervella_2004_lCar_paper,
       author = {{Kervella}, Pierre and {Fouqu{\'e}}, Pascal and {Storm}, Jesper and {Gieren}, Wolfgang P. and {Bersier}, David and {Mourard}, Denis and {Nardetto}, Nicolas and {du Coud{\'e} Foresto}, Vincent},
        title = "{The Angular Size of the Cepheid l Carinae: A Comparison of the Interferometric and Surface Brightness Techniques}",
      journal = {\apjl},
         year = 2004,
        month = apr,
       volume = {604},
       number = {2},
        pages = {L113-L116},
          doi = {10.1086/383571},
archivePrefix = {arXiv},
       eprint = {astro-ph/0402244},
 primaryClass = {astro-ph},
       adsurl = {https://ui.adsabs.harvard.edu/abs/2004ApJ...604L.113K}
}

@ARTICLE{perryman_1997_hypparcos,
       author = {{Perryman}, M.~A.~C. and {Lindegren}, L. and {Kovalevsky}, J. and {Hoeg}, E. and {Bastian}, U. and {Bernacca}, P.~L. and {Cr{\'e}z{\'e}}, M. and {Donati}, F. and {Grenon}, M. and {Grewing}, M. and {van Leeuwen}, F. and {van der Marel}, H. and {Mignard}, F. and {Murray}, C.~A. and {Le Poole}, R.~S. and {Schrijver}, H. and {Turon}, C. and {Arenou}, F. and {Froeschl{\'e}}, M. and {Petersen}, C.~S.},
        title = "{The HIPPARCOS Catalogue}",
      journal = {\aap},
         year = 1997,
        month = jul,
       volume = {323},
        pages = {L49-L52},
       adsurl = {https://ui.adsabs.harvard.edu/abs/1997A&A...323L..49P}
}

@ARTICLE{Leeuwen_2007_newHipparcosPllx,
       author = {{van Leeuwen}, F.},
        title = "{Validation of the new Hipparcos reduction}",
      journal = {\aap},
         year = 2007,
        month = nov,
       volume = {474},
       number = {2},
        pages = {653-664},
          doi = {10.1051/0004-6361:20078357},
archivePrefix = {arXiv},
       eprint = {0708.1752},
 primaryClass = {astro-ph},
       adsurl = {https://ui.adsabs.harvard.edu/abs/2007A&A...474..653V}
}

@ARTICLE{Leeuwen_2007_cepDistances,
       author = {{van Leeuwen}, Floor and {Feast}, Michael W. and {Whitelock}, Patricia A. and {Laney}, Clifton D.},
        title = "{Cepheid parallaxes and the Hubble constant}",
      journal = {\mnras},
         year = 2007,
        month = aug,
       volume = {379},
       number = {2},
        pages = {723-737},
          doi = {10.1111/j.1365-2966.2007.11972.x},
archivePrefix = {arXiv},
       eprint = {0705.1592},
 primaryClass = {astro-ph},
       adsurl = {https://ui.adsabs.harvard.edu/abs/2007MNRAS.379..723V}
}

@ARTICLE{benedict_2007,
       author = {{Benedict}, G. Fritz and {McArthur}, Barbara E. and {Feast}, Michael W. and {Barnes}, Thomas G. and {Harrison}, Thomas E. and {Patterson}, Richard J. and {Menzies}, John W. and {Bean}, Jacob L. and {Freedman}, Wendy L.},
        title = "{Hubble Space Telescope Fine Guidance Sensor Parallaxes of Galactic Cepheid Variable Stars: Period-Luminosity Relations}",
      journal = {\aj},
         year = 2007,
        month = apr,
       volume = {133},
       number = {4},
        pages = {1810-1827},
          doi = {10.1086/511980},
archivePrefix = {arXiv},
       eprint = {astro-ph/0612465},
 primaryClass = {astro-ph},
       adsurl = {https://ui.adsabs.harvard.edu/abs/2007AJ....133.1810B}
}

@ARTICLE{benedict_2002,
       author = {{Benedict}, G. Fritz and {McArthur}, B.~E. and {Fredrick}, L.~W. and {Harrison}, T.~E. and {Slesnick}, C.~L. and {Rhee}, J. and {Patterson}, R.~J. and {Skrutskie}, M.~F. and {Franz}, O.~G. and {Wasserman}, L.~H. and {Jefferys}, W.~H. and {Nelan}, E. and {van Altena}, W. and {Shelus}, P.~J. and {Hemenway}, P.~D. and {Duncombe}, R.~L. and {Story}, D. and {Whipple}, A.~L. and {Bradley}, A.~J.},
        title = "{Astrometry with the Hubble Space Telescope: A Parallax of the Fundamental Distance Calibrator {\ensuremath{\delta}} Cephei}",
      journal = {\aj},
         year = 2002,
        month = sep,
       volume = {124},
       number = {3},
        pages = {1695-1705},
          doi = {10.1086/342014},
archivePrefix = {arXiv},
       eprint = {astro-ph/0206214},
 primaryClass = {astro-ph},
       adsurl = {https://ui.adsabs.harvard.edu/abs/2002AJ....124.1695B}
}

@ARTICLE{lane_2002,
       author = {{Lane}, Benjamin F. and {Creech-Eakman}, Michelle J. and {Nordgren}, Tyler E.},
        title = "{Long-Baseline Interferometric Observations of Cepheids}",
      journal = {\apj},
         year = 2002,
        month = jul,
       volume = {573},
       number = {1},
        pages = {330-337},
          doi = {10.1086/340558},
archivePrefix = {arXiv},
       eprint = {astro-ph/0203060},
 primaryClass = {astro-ph},
       adsurl = {https://ui.adsabs.harvard.edu/abs/2002ApJ...573..330L}
}

@ARTICLE{wang_2023,
       author = {{Wang}, Shu and {Chen}, Xiaodian},
        title = "{The Optical to Infrared Extinction Law of Magellanic Clouds Based on Red Supergiants and Classical Cepheids}",
      journal = {\apj},
         year = 2023,
        month = mar,
       volume = {946},
       number = {1},
          eid = {43},
        pages = {43},
          doi = {10.3847/1538-4357/acb647},
archivePrefix = {arXiv},
       eprint = {2301.09146},
 primaryClass = {astro-ph.GA},
       adsurl = {https://ui.adsabs.harvard.edu/abs/2023ApJ...946...43W}
}

@ARTICLE{riess_2022,
       author = {{Riess}, Adam G. and {Yuan}, Wenlong and {Macri}, Lucas M. and {Scolnic}, Dan and {Brout}, Dillon and {Casertano}, Stefano and {Jones}, David O. and {Murakami}, Yukei and {Anand}, Gagandeep S. and {Breuval}, Louise and {Brink}, Thomas G. and {Filippenko}, Alexei V. and {Hoffmann}, Samantha and {Jha}, Saurabh W. and {D'arcy Kenworthy}, W. and {Mackenty}, John and {Stahl}, Benjamin E. and {Zheng}, WeiKang},
        title = "{A Comprehensive Measurement of the Local Value of the Hubble Constant with 1 km s$^{-1}$ Mpc$^{-1}$ Uncertainty from the Hubble Space Telescope and the SH0ES Team}",
      journal = {\apjl},
         year = 2022,
        month = jul,
       volume = {934},
       number = {1},
          eid = {L7},
        pages = {L7},
          doi = {10.3847/2041-8213/ac5c5b},
archivePrefix = {arXiv},
       eprint = {2112.04510},
 primaryClass = {astro-ph.CO},
       adsurl = {https://ui.adsabs.harvard.edu/abs/2022ApJ...934L...7R}
}

@ARTICLE{riess_2024,
       author = {{Riess}, Adam G. and {Scolnic}, Dan and {Anand}, Gagandeep S. and {Breuval}, Louise and {Casertano}, Stefano and {Macri}, Lucas M. and {Li}, Siyang and {Yuan}, Wenlong and {Huang}, Caroline D. and {Jha}, Saurabh and {Murakami}, Yukei S. and {Beaton}, Rachael and {Brout}, Dillon and {Wu}, Tianrui and {Addison}, Graeme E. and {Bennett}, Charles and {Anderson}, Richard I. and {Filippenko}, Alexei V. and {Carr}, Anthony},
        title = "{JWST Validates HST Distance Measurements: Selection of Supernova Subsample Explains Differences in JWST Estimates of Local H $_{0}$}",
      journal = {\apj},
         year = 2024,
        month = dec,
       volume = {977},
       number = {1},
          eid = {120},
        pages = {120},
          doi = {10.3847/1538-4357/ad8c21},
archivePrefix = {arXiv},
       eprint = {2408.11770},
 primaryClass = {astro-ph.CO},
       adsurl = {https://ui.adsabs.harvard.edu/abs/2024ApJ...977..120R}
}

@ARTICLE{CosmoVerse_valentino_2025,
       author = {{Di Valentino}, Eleonora and {Said}, Jackson Levi and {Riess}, Adam and {Pollo}, Agnieszka and {Poulin}, Vivian and {G{\'o}mez-Valent}, Adri{\`a} and {Weltman}, Amanda and {Palmese}, Antonella and {Huang}, Caroline D. and {van de Bruck}, Carsten and {Saraf}, Chandra Shekhar and {Kuo}, Cheng-Yu and {Uhlemann}, Cora and {Grand{\'o}n}, Daniela and {Paz}, Dante and {Eckert}, Dominique and {Teixeira}, Elsa M. and {Saridakis}, Emmanuel N. and {Colg{\'a}in}, Eoin {\'O}. and {Beutler}, Florian and {Niedermann}, Florian and {Bajardi}, Francesco and {Barenboim}, Gabriela and {Gubitosi}, Giulia and {Musella}, Ilaria and {Banik}, Indranil and {Szapudi}, Istvan and {Singal}, Jack and {Cases}, Jaume Haro and {Chluba}, Jens and {Torrado}, Jes{\'u}s and {Mifsud}, Jurgen and {Jedamzik}, Karsten and {Said}, Khaled and {Dialektopoulos}, Konstantinos and {Herold}, Laura and {Perivolaropoulos}, Leandros and {Zu}, Lei and {Galbany}, Llu{\'\i}s and {Breuval}, Louise and {Visinelli}, Luca and {Escamilla}, Luis A. and {Anchordoqui}, Luis A. and {Sheikh-Jabbari}, M.~M. and {Lembo}, Margherita and {Dainotti}, Maria Giovanna and {Vincenzi}, Maria and {Asgari}, Marika and {Gerbino}, Martina and {Forconi}, Matteo and {Cantiello}, Michele and {Moresco}, Michele and {Benetti}, Micol and {Sch{\"o}neberg}, Nils and {Akarsu}, {\"O}zg{\"u}r and {Nunes}, Rafael C. and {Bernardo}, Reginald Christian and {Ch{\'a}vez}, Ricardo and {Anderson}, Richard I. and {Watkins}, Richard and {Capozziello}, Salvatore and {Li}, Siyang and {Vagnozzi}, Sunny and {Pan}, Supriya and {Treu}, Tommaso and {Irsic}, Vid and {Handley}, Will and {Giar{\`e}}, William and {Murakami}, Yukei and {Banihashemi}, Abdolali and {Poudou}, Ad{\`e}le and {Heavens}, Alan and {Kogut}, Alan and {Domi}, Alba and {Lenart}, Aleksander {\L}ukasz and {Melchiorri}, Alessandro and {Vadal{\`a}}, Alessandro and {Amon}, Alexandra and {Rivera}, Alexander Bonilla and {Reeves}, Alexander and {Zhuk}, Alexander and {Bonanno}, Alfio and {{\"O}vg{\"u}n}, Ali and {Pisani}, Alice and {Talebian}, Alireza and {Abebe}, Amare and {Aboubrahim}, Amin and {Gonz{\'a}lez Mor{\'a}n}, Ana Luisa and {Kov{\'a}cs}, Andr{\'a}s and {Lymperis}, Andreas and {Papatriantafyllou}, Andreas and {Liddle}, Andrew R. and {Paliathanasis}, Andronikos and {Borowiec}, Andrzej and {Yadav}, Anil Kumar and {Yadav}, Anita and {Sen}, Anjan Ananda and {William}, Anjitha John and {Davis}, Anne Christine and {Shajib}, Anowar J. and {Walters}, Anthony and {Lonappan}, Anto Idicherian and {Chudaykin}, Anton and {Capodagli}, Antonio and {da Silva}, Antonio and {De Felice}, Antonio and {Racioppi}, Antonio and {Oficial}, Araceli Soler and {Montiel}, Ariadna and {Favale}, Arianna and {Bernui}, Armando and {Velasco}, Arrianne Crystal and {Heinesen}, Asta and {Bakopoulos}, Athanasios and {Chatzistavrakidis}, Athanasios and {Khanpour}, Bahman and {Sathyaprakash}, Bangalore S. and {Zgirski}, Bartek and {L'Huillier}, Benjamin and {Famaey}, Benoit and {Jain}, Bhuvnesh and {Zhang}, Bing and {Karmakar}, Biswajit and {Dragovich}, Branko and {Thomas}, Brooks and {Correa}, Carlos and {Boiza}, Carlos G. and {Marques}, Catarina and {Escamilla-Rivera}, Celia and {Tzerefos}, Charalampos and {Zhang}, Chi and {De Leo}, Chiara and {Pfeifer}, Christian and {Lee}, Christine and {Venter}, Christo and {Gomes}, Cl{\'a}udio and {Roque De bom}, Clecio and {Moreno-Pulido}, Cristian and {Iosifidis}, Damianos and {Grin}, Dan and {Blixt}, Daniel and {Scolnic}, Dan and {Oriti}, Daniele and {Dobrycheva}, Daria and {Bettoni}, Dario and {Benisty}, David and {Fern{\'a}ndez-Arenas}, David and {Wiltshire}, David L. and {Sanchez Cid}, David and {Tamayo}, David and {Valls-Gabaud}, David and {Pedrotti}, Davide and {Wang}, Deng and {Staicova}, Denitsa and {Totolou}, Despoina and {Rubiera-Garcia}, Diego and {Milakovi{\'c}}, Dinko and {Pesce}, Dominic W. and {Sluse}, Dominique and {Borka}, Du{\v{s}}ko and {Yusofi}, Ebrahim and {Giusarma}, Elena and {Terlevich}, Elena and {Tomasetti}, Elena and {Vagenas}, Elias C. and {Fazzari}, Elisa and {Ferreira}, Elisa G.~M. and {Barakovic}, Elvis and {Dimastrogiovanni}, Emanuela and {Holm}, Emil Brinch and {Mottola}, Emil and {{\"O}z{\"u}lker}, Emre and {Specogna}, Enrico and {Brocato}, Enzo and {Jensko}, Erik and {Enriquez}, Erika Antonette and {Bhatia}, Esha and {Bresolin}, Fabio and {Avila}, Felipe and {Bouch{\`e}}, Filippo and {Bombacigno}, Flavio and {Anagnostopoulos}, Fotios K. and {Pace}, Francesco and {Sorrenti}, Francesco and {Lobo}, Francisco S.~N. and {Courbin}, Fr{\'e}d{\'e}ric and {Hansen}, Frode K. and {Sloan}, Greg and {Farrugia}, Gabriel and {Lynch}, Gabriel and {Garcia-Arroyo}, Gabriela and {Raimondo}, Gabriella and {Lambiase}, Gaetano and {Anand}, Gagandeep S. and {Poulot}, Gaspard and {Leon}, Genly and {Kouniatalis}, Gerasimos and {Nardini}, Germano and {Cs{\"o}rnyei}, G{\'e}za and {Galloni}, Giacomo},
        title = "{The CosmoVerse White Paper: Addressing observational tensions in cosmology with systematics and fundamental physics}",
      journal = {Physics of the Dark Universe},
         year = 2025,
        month = sep,
       volume = {49},
          eid = {101965},
        pages = {101965},
          doi = {10.1016/j.dark.2025.101965},
archivePrefix = {arXiv},
       eprint = {2504.01669},
 primaryClass = {astro-ph.CO},
       adsurl = {https://ui.adsabs.harvard.edu/abs/2025PDU....4901965D}
}

@ARTICLE{valle_2024,
       author = {{Valle}, G. and {Dell'Omodarme}, M. and {Prada Moroni}, P.~G. and {Degl'Innocenti}, S.},
        title = "{Testing the asteroseismic estimates of stellar radii with surface brightness-colour relations and Gaia DR3 parallaxes: Red giants and red clump stars}",
      journal = {\aap},
         year = 2024,
        month = oct,
       volume = {690},
          eid = {A327},
        pages = {A327},
          doi = {10.1051/0004-6361/202451473},
archivePrefix = {arXiv},
       eprint = {2409.10050},
 primaryClass = {astro-ph.SR},
       adsurl = {https://ui.adsabs.harvard.edu/abs/2024A&A...690A.327V}
}

@ARTICLE{campante_2019,
       author = {{Campante}, Tiago L. and {Corsaro}, Enrico and {Lund}, Mikkel N. and {Mosser}, Beno{\^\i}t and {Serenelli}, Aldo and {Veras}, Dimitri and {Adibekyan}, Vardan and {Antia}, H.~M. and {Ball}, Warrick and {Basu}, Sarbani and {Bedding}, Timothy R. and {Bossini}, Diego and {Davies}, Guy R. and {Delgado Mena}, Elisa and {Garc{\'\i}a}, Rafael A. and {Handberg}, Rasmus and {Hon}, Marc and {Kane}, Stephen R. and {Kawaler}, Steven D. and {Kuszlewicz}, James S. and {Lucas}, Miles and {Mathur}, Savita and {Nardetto}, Nicolas and {Nielsen}, Martin B. and {Pinsonneault}, Marc H. and {Reffert}, Sabine and {Silva Aguirre}, V{\'\i}ctor and {Stassun}, Keivan G. and {Stello}, Dennis and {Stock}, Stephan and {Vrard}, Mathieu and {Y{\i}ld{\i}z}, Mutlu and {Chaplin}, William J. and {Huber}, Daniel and {Bean}, Jacob L. and {{\c{C}}elik Orhan}, Zeynep and {Cunha}, Margarida S. and {Christensen-Dalsgaard}, J{\o}rgen and {Kjeldsen}, Hans and {Metcalfe}, Travis S. and {Miglio}, Andrea and {Monteiro}, M{\'a}rio J.~P.~F.~G. and {Nsamba}, Benard and {{\"O}rtel}, Sibel and {Pereira}, Filipe and {Sousa}, S{\'e}rgio G. and {Tsantaki}, Maria and {Turnbull}, Margaret C.},
        title = "{TESS Asteroseismology of the Known Red-giant Host Stars HD 212771 and HD 203949}",
      journal = {\apj},
         year = 2019,
        month = nov,
       volume = {885},
       number = {1},
          eid = {31},
        pages = {31},
          doi = {10.3847/1538-4357/ab44a8},
archivePrefix = {arXiv},
       eprint = {1909.05961},
 primaryClass = {astro-ph.SR},
       adsurl = {https://ui.adsabs.harvard.edu/abs/2019ApJ...885...31C}
}

@ARTICLE{storm_2011_A94_I,
       author = {{Storm}, J. and {Gieren}, W. and {Fouqu{\'e}}, P. and {Barnes}, T.~G. and {Pietrzy{\'n}ski}, G. and {Nardetto}, N. and {Weber}, M. and {Granzer}, T. and {Strassmeier}, K.~G.},
        title = "{Calibrating the Cepheid period-luminosity relation from the infrared surface brightness technique. I. The p-factor, the Milky Way relations, and a universal K-band relation}",
      journal = {\aap},
         year = 2011,
        month = oct,
       volume = {534},
          eid = {A94},
        pages = {A94},
          doi = {10.1051/0004-6361/201117155},
archivePrefix = {arXiv},
       eprint = {1109.2017},
 primaryClass = {astro-ph.CO},
       adsurl = {https://ui.adsabs.harvard.edu/abs/2011A&A...534A..94S}
}

\begin{appendix}
\end{appendix}
\end{document}